\documentclass[a4paper,11pt]{article}
\pdfoutput=1
 \usepackage{jheppub} 
 \usepackage{graphicx}
\usepackage{amsmath}
 \usepackage{amssymb}
 \usepackage{slashed}
\usepackage{dcolumn}% Align table columns on decimal point
\usepackage{bm}% bold math 
\usepackage{color}
\usepackage{multirow}
\usepackage{comment,ulem}

\allowdisplaybreaks

 \newcommand{\lsim}{{\;\raise0.3ex\hbox{$<$\kern-0.75em\raise-1.1ex\hbox{$\sim$}}\;}}
\newcommand{\gsim}{{\;\raise0.3ex\hbox{$>$\kern-0.75em\raise-1.1ex\hbox{$\sim$}}\;}}
\def\bea{\begin{eqnarray}}
\def\eea{\end{eqnarray}}
\def\bec{\begin{center}}
\def\ec{\end{center}}

\def\beq{\begin{equation}}
\def\eeq{\end{equation}}

\def\bea{\begin{eqnarray}}
\def\eea{\end{eqnarray}}
\def\beq#1\eeq{\begin{align}#1\end{align}}
\def\beqnn#1\eeq{\begin{align*}#1\end{align*}}
\def\ba{\begin{array}}
\def\ea{\end{array}}
\def\bc{\begin{center}}
\def\ec{\end{center}}

\preprint{CTPU-PTC-21-35}

\title{Axion emission from supernova with axion-pion-nucleon contact interaction}
 
\author[a]{Kiwoon Choi,}
\author[a]{Hee Jung Kim,}
\author[a]{Hyeonseok Seong}
\author[b,a]{and Chang Sub Shin}

\affiliation[a]{Center for Theoretical Physics of the Universe, Institute for Basic Science (IBS),\\
  34126 Daejeon, South Korea }
  
\affiliation[b]{Department of Physics and Institute of Quantum Systems (IQS), Chungnam National University,\\
34134 Daejeon, South Korea}

\emailAdd{kchoi@ibs.re.kr}
\emailAdd{heejungkim@ibs.re.kr}
\emailAdd{hseong@ibs.re.kr}
\emailAdd{csshin@cnu.ac.kr}

\abstract{
We examine the axion emission from  supernovae with a complete set of relevant axion couplings 
including the axion-pion-nucleon contact interaction which was ignored in the previous studies. 
Two processes are affected by the axion-pion-nucleon contact interaction, 
$\pi^-+p \rightarrow n + a$ 
and $n+p\rightarrow n+p+a$, and these processes can be the dominant source of axions for some region in the axion parameter space or in astrophysical conditions encountered inside supernovae.
We find that the contact interaction can enhance the axion emissivity of $\pi^-+p \rightarrow n + a$
by a factor of $2-4$, while the effect on $n+p\rightarrow n+p+a$ is not significant. 
We also discuss the relative importance of other pion-induced processes 
such as $\pi^0+n\rightarrow n+a$ and $\pi^-+\pi^0\rightarrow \pi^-+a$.}
\vspace{4cm}

\begin{document} 
\maketitle
\flushbottom
 
\section{Introduction}

The axion which was initially introduced as a solution to the strong CP problem   \cite{Peccei:1977hh, Peccei:1977ur, Weinberg:1977ma, Wilczek:1977pj}  has turned out to have many interesting phenomenological consequences \cite{Kim:2008hd,DiLuzio:2020wdo,Choi:2020rgn}.
%
%Since the realization that axions provide a compelling candidate for dark matter,
After recognizing that axions provide a compelling candidate for the dark matter in the Universe \cite{Preskill:1982cy,Abbott:1982af,Dine:1982ah}, 
%especially for the axion mass $m_a = \mathcal{O}(\mu {\rm eV} - {\rm meV})$, 
a lot of efforts have been made to search for axions over 
the parameter space of the representative axion models \cite{Kim:1979if,Shifman:1979if,Dine:1981rt,Zhitnitsky:1980tq}. 
Since the viable parameter region is in the very weakly coupled regime, most of the laboratory experiments searching for axions are ongoing in the direction of the 
precision measurements using, for example, resonant cavities, 
%\cite{Sikivie:1983ip, Khatiwada:2020mld, Semertzidis:2019gkj} 
nuclear magnetic resonance, 
%\cite{JacksonKimball:2017elr}, 
light shining through the walls, 
%\cite{Bahre:2013ywa}, 
and polarization of lights in magnetic fields (see \cite{Graham:2015ouw,Irastorza:2018dyq,Semertzidis:2021rxs} for comprehensive reviews).
%\cite{Ejlli:2020yhk}. 
%
%Although it is not an easy task to confirm the existence of the axion in laboratories, the constant %endeavors make slow but steady progress.
%

%
A complementary approach which can severely constrain the couplings of light axions is to use
astrophysical objects forming a hot and dense environment,
e.g., supernovae, stars on the horizontal and red giant branches, neutron stars, and even white dwarfs
(see \cite{Raffelt:2006cw}  for a review and also \cite{DiLuzio:2021ysg} for a recent overview).
Axions can be produced abundantly from those objects, thereby altering their evolution.
%
%With this consideration, one might be able to  
%Depending on how precisely we understand  the corresponding astrophysical object, 
One can then derive constraints on the couplings of
 axions by requiring that the axion emission does not significantly alter the standard evolution scenario which is consistent with the observational data.

A core-collapse supernova, e.g., SN1987A, is known to provide stringent constraints on the 
axion couplings to hadrons, particularly on the couplings to nucleons \cite{Turner:1987by,Raffelt:1987yt}.
% one of the hottest and abrupt phenomena we have observed. 
The observation of the neutrino flux from  SN1987A, 
%the neutrino emission lasted over $\sim10$ seconds 
which is consistent with the standard scenario \cite{Burrows:2000mk,Woosley:2005cha},
%The further observations of the event during 30 years provide strong evidence 
%of the existence of the neutron star as the remnant of SN1987A \cite{Cigan:2019shp,Page:2020gsx}.
%
suggests that the additional cooling by axion emission from the associated proto-neutron star 
is constrained as $L_a \lesssim L_\nu = {\cal O}(1- 10) \times 10^{51}\,{\rm erg}/\sec$, where $L_a$ and $L_\nu$ denote the axion  and neutrino luminosities  around $1-10\,\sec$ after the formation of the proto-neutron star \cite{Raffelt:2006cw}.
%at a temperature $T\sim 30\,$MeV with a mass density $\rho\sim 2\times 10^{14}\, {\rm g}/{\rm cm}^3$  
%If the additional energy loss occurs, the duration would be shorter, and the extra energy loss 
%would be constrained with an order of magnitude estimation. 
%In the presence of the axion, the additional cooling of a proto-neutron star can be supported by the %axion emission channels besides the neutrino emission.
%Then, for the axion luminosity 
%\bea 
%L_a = \int d^3{\bf r} \sum_i Q_a^i, 
%\eea 
%where $Q_a^i$ is the axion emissivity for a specific axion production channel $i$, 
%we can constrain the axion parameters by 
%
%If we consider the energy loss through the axion couplings to nucleon, pion, lepton, and even the gauge bosons, we could constrain the parameter space of them. 
%

Among the processes  producing axions from supernovae, the nucleon bremsstrahlung process $N+N\rightarrow N+N+a$ ($N=n,p$) has been considered as the dominating process 
 for many years \cite{Iwamoto:1984ir,Brinkmann:1988vi,Raffelt:1993ix,Iwamoto:1992jp,Carenza:2019pxu}.
However, recently it has been noticed that the number density of negatively charged pions inside supernovae can be significantly enhanced by pion-nucleon interactions \cite{Fore:2019wib}.
Based on this observation, the pion-induced Compton-like process $\pi^-+p \rightarrow n + a$, which was originally studied in \cite{Turner:1991ax,Keil:1996ju}, has been revisited. Taking into account medium effects, Refs.~\cite{Carenza:2020cis,Fischer:2021jfm} show that the process dominates over the nucleon-nucleon bremsstrahlung for a wide range of astrophysical conditions encountered inside supernovae.\footnote{
The medium effects also modify the axion-nucleon couplings. The modification is expected to be an O(1) effect in general, while it could result in $\sim$10-times enhancement of the axion-neutron coupling in the KSVZ model because the accidental cancellation of that coupling in vacuum is spoiled \cite{Balkin:2020dsr}. Here we presume the values of axion couplings in vacuum for our numerical estimation.
}
 % The medium effects also modify the axion-nucleon couplings. The modification of the couplings could be $\mathcal{O}(1)$. In particular, the axion-neutron coupling in the KSVZ model can be enhanced by a factor of $\sim$10; this is because the accidental cancellation of that coupling in vacuum is spoiled by the medium effects \cite{Balkin:2020dsr}. Here we presume the values of axion couplings in vacuum for numerical estimation.

 Motivated by the importance of the process $\pi^-+p \rightarrow n + a$, in this paper we extend the previous analysis of axion emission
from supernovae  with a complete set of relevant axion couplings including the axion-pion-nucleon and axion-pion {\it contact interactions} which were ignored in the previous studies. 
Our primary concern is how significantly the contact interactions can affect the axion emissivity. 
We start with a general axion Lagrangian above the QCD confinement scale, which determines the axion couplings to hadrons below that scale.
%We examine the effective axion couplings for some representative axion models.
To highlight the coupling dependence of the axion emissivity more clearly, we compare a new contribution including the effect of contact terms to that from the axion-nucleon couplings only and take the ratio between the two contributions. It is expected that the ratios can lead to cancellation of the uncertainties in nuclear physics and the medium effect. Thus, as a first step towards understanding the contributions of the contact interactions, we consider the tree-level diagrams in the leading order pion-nucleon couplings and the one-pion exchange diagrams for the nucleon-nucleon bremsstrahlung. We also ignore the background matter effect, which should be included in future work.
%In this way, we separate the part that depends on the axion couplings maximally from the phase space integration that depends more on the nulear physics and the  environmental parameter.  
%the relative importance of each term becomes clear.
In such an approximation, two processes are affected by the axion-pion-nucleon contact interaction,  $\pi^-+p \rightarrow n + a$ and $n+p\rightarrow
n+p+a$.   We find   
that the axion-pion-nucleon contact interaction can enlarge
the emission rate  of $\pi^-+p \rightarrow n + a$
by a factor of $2-4$ depending on the pattern of axion couplings,  while the effect on $n+p\rightarrow
n+p+a$ is negligible.
We also examine other  pion-induced processes such as $\pi^0+ n\rightarrow n+a$ and $\pi^-+\pi^0\rightarrow \pi^-+a$, where the latter process is induced by the axion-pion contact interaction. We then find that $\pi^0+ n\rightarrow n+a$ can be as important as $\pi^-+p \rightarrow n + a$, again depending on the pattern of axion couplings, while   $\pi^-+\pi^0\rightarrow \pi^-+a$ is negligible compared to $\pi^-+p \rightarrow n + a$ over the entire axion parameter space for astrophysical conditions encountered inside proto-neutron stars.

%
%The importance of each process depends on  the ambient conditions inside supernovae and also on the %pattern of  axion couplings  which can be derived from the underlying UV models.

% such as the KSVZ  model, DFSZ model, and 
%other variations \cite{}. 
%
%We consider the coupling dependence in detail and identify the parameter region where the extra %contributions discussed here could be maximized.
%

This paper is organized as follows.
In Sec.~\ref{Sec:UVmodel}, we introduce the relevant axion couplings to nucleons and pions in the context of a generic axion model and discuss the model dependence of couplings for a simple class of axion models.
In Sec.~\ref{Sec:Calc}, we investigate the axion emission from supernovae by a variety of pion-induced processes and the nucleon-nucleon bremsstrahlung processes, with a complete set of relevant axion couplings.
% including the axion-pion-nucleon contact interaction.
%The other axion emission channels, which could be relevant in potential, are discussed as well.
%
Sec.~\ref{Sec:Conc} is a summary and conclusion.

\section{Axion Couplings to Nucleons and Pions \label{Sec:UVmodel}}

%Since the couplings between the axion and hadrons are most relevant for the axion emission in the supernova explosion, 
In this section, we briefly discuss the axion couplings to nucleons and pions for generic axions whose  couplings are constrained only by the (approximate) global $U(1)$ Peccei-Quinn (PQ) symmetry
\cite{Peccei:1977hh, Peccei:1977ur, Weinberg:1977ma, Wilczek:1977pj}.
% the gluons and the light quarks $u$ and $d$.
%Although we are focuing on QCD axions, our discussion is also applicable to more general axion-like %particles (ALPs). The only assumption we make on the underlying model is the existence  of a global %Peccei-Quinn (PQ) $U(1)$ symmetry which is explicitly broken dominantly by the QCD anomaly, which %would assure that the strong CP problem is solved by the axion mechanism 
Without loss of generality, at scales below the axion decay constant $f_a$, one can always choose a field basis for which {\it only} the axion field transforms under  the PQ symmetry as
\bea 
U(1)_{\rm PQ}:\quad  a \, \to \,a+ {\rm constant},
\eea 
while all other fields are invariant  \cite{Georgi:1986df}. 
In such a field basis, the axion couplings  at low energy scales around $\mu={\cal O}(1)$ GeV include
 \bea \label{Lquark}
{\cal L}_{\rm eff} &=& 
 c_G\frac{g_s^2}{32\pi^2}\frac{a}{f_a} G_{\mu\nu}^a \tilde G^{a\mu\nu} 
 + \frac{\partial_\mu a}{2 f_a} \Big( C_u \bar u \gamma^\mu \gamma_5 u 
 + C_d\bar d \gamma^\mu\gamma_5 d \Big),
\eea 
where the axion decay constant $f_a$ defines the axion field range as
$a\cong a+2\pi f_a$, $G^a_{\mu\nu}$ are the gluon field strength, and $u$ and $d$ are the up and down quarks. Here $c_G$ is an integer-valued  parameter describing the $U(1)_{\rm PQ}$ breaking by the QCD anomaly, while $C_u$ and $C_d$ are continuous real-valued parameters describing  the
$U(1)_{\rm PQ}$-preserving  axion couplings to the light quarks  renormalized at  $\mu={\cal O}(1)$ GeV.

For axion models which have a UV completion with a linearly realized $U(1)_{\rm PQ}$,
the low energy parameters $c_G$ and  $C_{u,d}$ in Eq.~\eqref{Lquark} are determined mainly by the $U(1)_{\rm PQ}$ charges defined 
in the UV model.\footnote{
For string-theoretic axions that arise from the zero modes of  higher-dimensional $p$-form gauge field, there is no UV completion with a linearly realized $U(1)_{\rm PQ}$. It has been noted that the tree-level values of $C_{u,d}$ for string-theoretic axions
are of the order of $\alpha_{\rm GUT}/2\pi$
%which are numerically similar to those of the KSVZ axions
 \cite{Choi:2021kuy}. 
}
As an illustrative example, let us consider axion models
in which the first generation quark masses are generated by the following Yukawa couplings\footnote{Here for simplicity we ignore the effects of flavor mixings.}:
\bea 
\label{yukawa}
{\cal L}_{\rm Yukawa} 
=   \lambda_u \left(\frac{\sigma}{\Lambda}\right)^{n_u}Q_1 u^c_1 H_u 
+ \lambda_d \left(\frac{\sigma}{\Lambda}\right)^{n_d} Q_1 d^c_1 H_{d} + h.c.,
\eea 
where $\sigma$ is a PQ-charged gauge-singlet scalar field whose vacuum expectation value determines the axion decay constant as 
\bea
\langle \sigma\rangle =\frac{1}{\sqrt{2}}f_a e^{ia/f_a},\eea 
$Q_i$ and $u^c_i, d^c_i$ ($i=1,2,3$) denote the three generations of the left-handed $SU(2)_L$-doublet quarks and
the left-handed $SU(2)_L$-singlet antiquarks, respectively, $H_u$ and $H_d$ are  $SU(2)_L$-doublet Higgs fields, and finally $\Lambda$ is a cutoff scale of the model. 
To derive the low energy axion couplings in this model,
  we first make the following axion-dependent field redefinition at a scale around
 $f_a$:
 \bea
 \label{redef}
 \Phi\,\rightarrow \, e^{iq_\Phi a/f_a} \Phi\quad (\Phi=\psi, H_{u,d}),\eea
 and subsequently integrate out all massive fields heavier than $\mu={\cal O}(1)$ GeV,
 where  $q_\Phi$ is the PQ charge of $\Phi$ 
 (in the normalization convention with  $q_\sigma=1$) for the linearly realized 
 $U(1)_{\rm PQ}$, and 
 $\psi$ stands for all chiral fermions in the model.
Then the axion-gluon coupling $c_G$, which arises as a consequence of the axion-dependent field redefinition of $\psi$,   corresponds to the coefficient of the $U(1)_{\rm PQ}$-$SU(3)_c$-$SU(3)_c$ anomaly, while
the couplings $C_{u,d}$ to the light quarks are determined by (i) a contribution from the axion-dependent field-redefinition of $\{Q_1,u^c_1,d^c_1\}$, (ii) the tree-level threshold correction from the axion mixing with the $Z$ boson which is induced by the field redefinition of $H_{u,d}$, and finally (iii) the radiative corrections caused by the gauge and Yukawa couplings in the model
 \cite{Choi:2021kuy}. Putting these together,
 one finds
\bea
\label{cucd}
c_G& =&2\sum_\psi q_\psi {\rm Tr}(T^2_c(\psi)),\nonumber \\
C_u &=& -n_u - (q_{H_u}+q_{H_d})\cos^2\beta+\Delta C_u,\nonumber \\
C_d&=& -n_d  -(q_{H_u}+q_{H_d})\sin^2\beta+\Delta C_d,
\eea 
where $T_c(\psi)$ is the color charge of $\psi$,
 $\tan\beta =\langle H_u\rangle/\langle H_d\rangle$, and   the radiative corrections $\Delta C_{u,d}={\cal O}(10^{-2}-10^{-3})$ can be safely ignored if the tree level values of $C_{u,d}$ are of order unity \cite{Choi:2021kuy}. 

The above results indicate that a variety of different patterns of  $c_G$ and $C_{u,d}$ are possible even within the framework of relatively simple axion models. 
Let us present explicitly the parameter values for some examples. 
In the KSVZ model \cite{Kim:1979if,Shifman:1979if},  $H_u=(i\sigma_2 H_d)^*$ and all SM fields are neutral under the linearly realized $U(1)_{\rm PQ}$, and therefore $n_u=n_d=q_{H_u}=q_{H_d}=0$. The model also involves a heavy PQ-charged exotic quark ${\cal Q}$ generating  the $U(1)_{\rm PQ}$-$SU(3)_c$-$SU(3)_c$ with $c_G=1$.
The resulting couplings of the KSVZ axion at $\mu={\cal O}(1)$ GeV  are given by \bea
\label{model:KSVZ}
{\rm KSVZ:} \quad c_G=1, \quad  C_u=\Delta C_u={\cal O}(10^{-2}), \quad C_d=\Delta C_d ={\cal O}(10^{-2}),\eea
where $\Delta C_{u,d}$ are induced mostly by the axion-gluon coupling $c_G$ causing a running of $C_{u,d}$ over the scales from 
the mass of the exotic quark ${\cal Q}$  to $\mu={\cal O}(1)$ GeV \cite{Choi:2021kuy}.
On the other hand,
the minimal DFSZ model \cite{Dine:1981rt,Zhitnitsky:1980tq} has $n_u=n_d=0$, $q_{H_u}=q_{H_d}=-1$ and all chiral fermions in the SM model have
$q_\psi=1/2$, which result in \bea
\label{model:DFSZ}
{\rm DFSZ:}\quad c_G=6,\quad  C_u=2\cos^2\beta +\Delta C_u, \quad  C_d=2\sin^2\beta+\Delta C_d\eea
with  $\Delta C_{u,d}={\cal O}(10^{-3})$ which are smaller than those of the KSVZ model  because in the DFSZ model the running of $C_{u,d}$ starts from a lower scale around the top quark mass \cite{Choi:2021kuy}.
It is an interesting possibility that $U(1)_{\rm PQ}$ plays the role of a flavor symmetry
which explains the fermion mass hierarchies~\cite{Ema:2016ops,Calibbi:2016hwq,Bjorkeroth:2017tsz}.
%particularly explain why the light quark masses are much lighter than the weak scale \cite{},
In such a case,  $n_{u,d}$ can be non-zero integers and  the model can have a more diverse pattern of $c_G$ and $C_{u,d}$.
%Note that although $n_{u,d}$ in the Yukawa couplings (Eq.~\eqref{yukawa}) are required to be  non-negative, $n_{u,d}$ in the axion couplings (Eq.~\eqref{cucd}) can have negative values when $\sigma$ in Eq.~\eqref{yukawa} is replaced  with $\sigma^*$.
Note that, while $n_{u,d}$ in the Yukawa couplings (Eq.~\eqref{yukawa}) are required to be non-negative, the sign in front of $n_{u,d}$ in Eq.~\eqref{cucd} can be flipped by replacing $\sigma$ in Eq.~\eqref{yukawa} with $\sigma^*$.
%one can consider a case with $\sigma^\ast$ instead of $\sigma$ in Eq.~\eqref{yukawa};
%in such a case, one should replace $-n_{u,d}$ with $n_{u,d}$ in Eq.~\eqref{cucd}.
%Note that the axion couplings (Eq.~\eqref{cucd}) change their sign when we replace $\sigma$ in Eq.~\eqref{yukawa} with $\sigma^\ast$.
One can further generalize the model by introducing additional  $U(1)_{\rm PQ}$-charged Higgs doublet, and then $C_{u,d}$ receive additional contribution depending on the vacuum expectation value of the added Higgs field. With this observation, in the following we regard $C_u$ and $C_d$ as real-valued free parameters, and $c_G$ as an integer-valued additional free parameter, without specifying the underlying UV model.

%In the UV perspective of field-theoretic axion models,  nonzero $c_{u,d}^0$ can be obtained if the SM %particles are also charged under the $U(1)_{\rm PQ}$. 
%Considering a set of Higgs doublet fields $\{H_i\}$ with the same gauge charge as the SM Higgs %doublet, $H_{i_u}$ and $H_{i_d}$ denote the Higgs doublets coupled to $u$ and $d$ quarks, %respectively. 
%In a minimal Higgs scenario like the Standard Model, $H_{i_u}=H_{i_d} = H$, 
%while in the two-Higgs-doublet model of type II, 
%they are different as $H_{i_u} = H_u$, $H_{i_d}= H_d$. 
%Then, the general form of PQ symmetric Yukawa interactions for the light quarks is

%For the PQ charges of the Higgs doublets $\{q_{H_i}\}$ that yield the transformation
%\bea 
%U(1)_{\rm PQ}: H_i\to e^{ i q_{H_i} \alpha} H_i,
%\eea  the general expressions of $c_u^0$ and $c_d^0$ are given by 

%where $v^2 = \sum_i v_i^2$, 
%$v_i = \sqrt{2}\langle H_i\rangle$.  
%For the KSVZ axion model in which all charges $n_{u, d}$ and $q_{H_n}$ are vanishing, 
%$c_u^0=c_d^0=0$. For the DFSZ axion model, we have $n_{u,d}=0$, $q_{H_u}= - q_{H_d} = -1$, and 
%$N_{\rm DW}= 6$. 
%This results in $c_u^0 = v_d^2/3v^2= \cos^2(\beta)/3$, $c_d^2= v_u^2/3v^2=\sin^2(\beta)/3$.

From the couplings in Eq.~(\ref{Lquark}) defined at $\mu={\cal O}(1)$ GeV, we can derive the axion couplings to nucleons and pions which are relevant for the axion emission from supernova. 
Including the conventional pion-nucleon couplings, the interactions are given by \cite{Chang:1993gm,DiLuzio:2020wdo}
\bea \label{Lhadron}
{\cal L}_{\rm int}  &=& \frac{g_A}{2f_\pi} \left( \partial_\mu\pi^0 (\bar p\gamma^\mu\gamma_5 p - \bar n \gamma^\mu \gamma_5 n) 
+ \sqrt{2}\partial_\mu\pi^+\bar p \gamma^\mu \gamma_5 n 
+ \sqrt{2}\partial_\mu\pi^- \bar n \gamma^\mu\gamma_5 p \right)     \nonumber\\
 && + \frac{\partial_\mu a}{2 f_a} \left( C_{ap} \bar p \gamma^\mu\gamma_5 p 
 + C_{an} \bar n \gamma^\mu \gamma_5 n 
  + \frac{C_{a\pi N}}{f_\pi} ( i \pi^+ \bar p \gamma^\mu n - i \pi^- \bar n\gamma^\mu p)  \right) \nonumber\\
 &&+ \frac{\partial_\mu a}{2 f_a}   \frac{C_{a\pi}}{f_\pi} \Big( \pi^0 \pi^+ \partial^\mu\pi^- +\pi^0\pi^-\partial^\mu\pi^+ - 2\pi^+\pi^-\partial^\mu\pi^0 \Big),
\eea 
where $f_\pi= 92.4$ MeV is the pion decay constant and
\bea \label{Coeffs}
%g_A  &=& \Delta u - \Delta d,  
%g_0^{ud} = \Delta u + \Delta d, 
%\nonumber\\
C_{ap}- C_{an} &=& g_A \left(C_u - C_d + \Big(\frac{m_u - m_d}{m_u+ m_d} \Big)c_G\right),\nonumber\\
C_{ap} + C_{an} &=& g_0 \Big(C_u + C_d - c_G \Big), \nonumber\\
 C_{a\pi N} &=&  \frac{C_{ap} - C_{an}}{\sqrt{2} g_A}, \quad C_{a\pi} = \frac{2(C_{ap} -C_{an})}{3 g_A}
\eea 
with the nucleon matrix elements of the light quark axial vector currents given by
\bea
\label{n_matrix}
g_A&=&\Delta u -\Delta d \simeq 1.2723(23), \nonumber \\
g_0&=&\Delta u+\Delta d \simeq 0.521(53),\eea
where $s^\mu \Delta q=\langle N|\bar q\gamma^\mu\gamma_5 q|N\rangle$ ($q=u,d$) for the nucleon spin four vector $s^\mu$. Here the numerical value of $g_0$ is chosen for the axion-quark couplings
$C_{u,d}$ renormalized at 
$\mu=2$ GeV in the $\overline{\rm MS}$ scheme \cite{diCortona:2015ldu}, and the small contributions from the axion couplings to the heavier quarks $Q=\{s,c,b,t\}$ are ignored.

%In this expression, we ignore the contributions from sea quarks. 
%The matrix elements $\Delta u$, $\Delta d$ are defined as 
%$s^\mu \Delta f \equiv \langle p| f \gamma^\mu \gamma_5 |p\rangle$,  
%where $s^\mu$ is the spin vector, and $|p\rangle$ is the proton state at rest. 
%For the matching scale, $Q=2$ GeV in the $\overline{\rm MS}$ scheme, the matrix elements and the 
%ratio of the current quark masses are evaluated as  
%\bea 
% g_0 = 0.521(53),\quad g_A = 1.2723(23) ,\quad \frac{m_u}{m_d} = 0.48(3).
%\eea 

The above results show that the entire axion couplings to nucleons and pions, including the axion-pion-nucleon contact interaction $C_{a\pi N}$ and the axion-pion contact interaction $C_{a\pi}$, are determined by the two free parameters $C_{an}$ and $C_{ap}$. An interesting feature of these parameters is that 
in some axion models they can have a hierarchical pattern such as
$|C_{ap}|\gg |C_{an}|$ or $|C_{ap}-C_{an}|\gg |C_{ap}+C_{an}|$ without fine tuning of any continuous parameter in the underlying UV model. 
For instance, including the radiative corrections induced by the axion-gluon coupling $c_G$,
the KSVZ and string-theoretic axions have $|C_{u,d}|={\cal O}(10^{-2} c_G)$ \cite{Choi:2021kuy}, which results in \bea
|C_{ap}|\simeq 0.48 |c_G|\,\gg\, |C_{an}| = {\cal O}(10^{-2}|c_G|)\eea
 for the nucleon matrix elements in Eq.~\eqref{n_matrix} and the
light quark mass ratio $m_u/m_d=0.48(3)$. Also, for the axion couplings in Eq.~\eqref{cucd}, the anomaly coefficient $c_G$ and the tree level value of $C_u+C_d$ are all quantized parameters. Then, for a model with  $U(1)_{\rm PQ}$-charges yielding
\bea
c_G=-\left(n_u+n_d+q_{H_u}+q_{H_d}\right)={\cal O}(1)\,,
\label{eq:maximalcase}
\eea
the model predicts 
\bea
C_{ap}-C_{an}={\cal O}(1),\quad   C_{ap}+C_{an}=g_0(\Delta C_u+\Delta C_d)\lesssim{\cal O}(10^{-2})\,.
\eea

At any rate, axions generically  have  the axion-pion-nucleon contact interaction given by 
 $C_{a\pi N}=(C_{ap}-C_{an})/\sqrt{2}g_A$ ($g_A\simeq 1.27$) and the axion-pion  contact interaction  $C_{a\pi}=2(C_{ap}-C_{an})/3g_A$. On the other hand, these contact interactions  were not taken into account
 in the previous studies of axion emission from supernovae.
% the effects of the axion-pion-nucleon contact interaction 
%($C_{a\pi N}$) and the axion contact interaction to three pions ($C_{a\pi}$) were not considered. 
In Sec.~\ref{Sec:Calc}, we will examine the effects of those contact interactions on the axion emission rates to see how important they can be.

\section{Axion Emission from Supernovae by hadronic processes\label{Sec:Calc}}

In this section, we examine the axion production by hadron collisions inside a newly born proto-neutron star. We consider three types of processes,
the pion-nucleon scattering 
$\pi+N\rightarrow N+a$ ($N=n,p$), the nucleon-nucleon  bremsstrahlung   $N+N\rightarrow N+N+a$, and the pion-pion scattering $\pi+\pi\rightarrow \pi+a$.
The relative importance of each process depends on the pattern of axion couplings, as well as on
the density and temperature of the corresponding astrophysical environment.
Our prime goal is to examine the effects of the two contact interactions, the axion-pion-nucleon contact coupling $C_{a\pi N}$ and the axion-pion contact coupling $C_{a\pi}$ in Eq.~(\ref{Lhadron}), which were not taken into account before except for the nucleon-nucleon bremsstrahlung \cite{Carena:1988kr}. We will examine this question in a simple approximation keeping only the leading order in pion-nucleon couplings and ignoring medium effects.
Accordingly, we can find a simple form of the coupling dependence in that approximation, and it shows the relative importance of the contribution from each coupling at a rough estimate.

\subsection{Pion-nucleon scattering}

Let us first discuss the pion-nucleon scattering process
$\pi+N\rightarrow N+a$.
For  $T\sim 40$ MeV and the nucleon mass density $\rho\sim 10^{14} \,{\rm g}/{\rm cm}^3$
% number density $n_N\sim 0.16\, {\rm fm}^{-3}$ (the nuclear saturation density) 
encountered inside a proto-neutron star, the pion and nucleon number densities roughly obey~\cite{Fore:2019wib}
\bea
\frac{n_{\pi^0}}{n_{\pi^-}}\sim \frac{n_{\pi^+}}{n_{\pi^0}}\sim \frac{n_p}{n_n}= {\cal O}(0.1)\,.
\eea
It is then expected that $\pi^-+p\rightarrow n+a$ and $\pi^0+n\rightarrow  n+a$ are the dominating process depending upon the involved axion couplings. 
\begin{figure}[t]
\centering
\includegraphics[width=\textwidth]{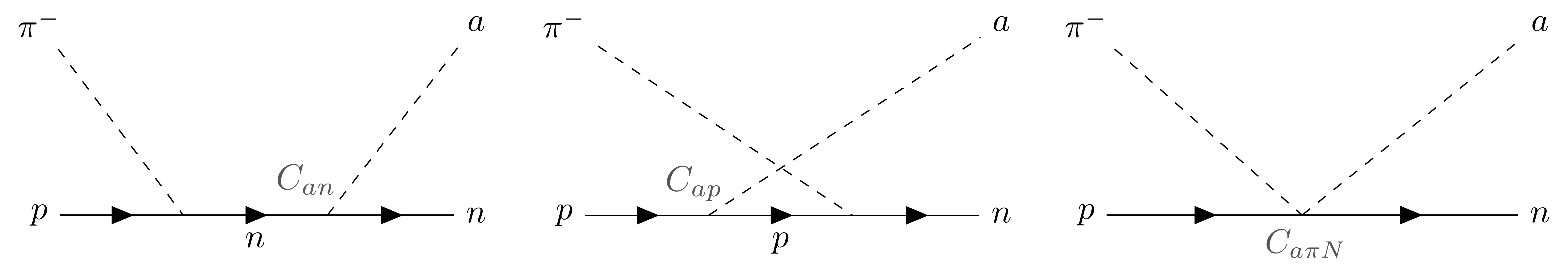} 
  \caption{Diagrams for $\pi^-+p\to n+a$ from the axion couplings in Eq.~(\ref{Lhadron}).}
\label{diagram:pionproton}
\end{figure}
\begin{figure}[t]
\centering
\includegraphics[width=0.65\textwidth]{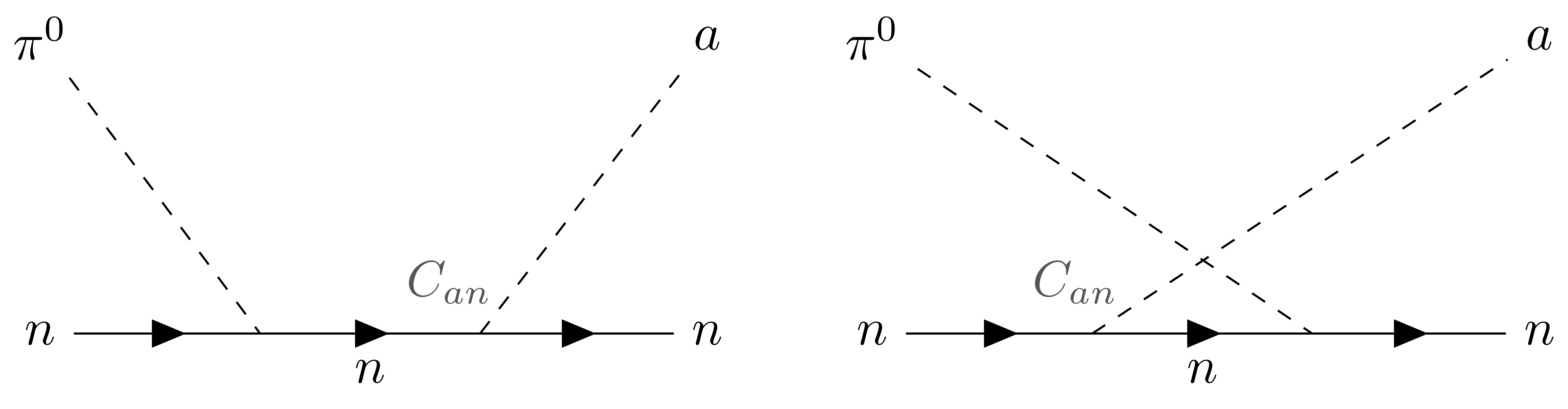} 
  \caption{Diagrams of axion production for the process $\pi^0+ n\to n+ a$.}
\label{diagram:npi0na}
\end{figure} 
The Feynman diagrams for these processes are depicted
in Fig.~\ref{diagram:pionproton} and Fig.~\ref{diagram:npi0na}, showing that at leading order in pion-nucleon couplings $\pi^0+n\rightarrow  n+a$ involves only the axion-neutron coupling $C_{an}$, while
$\pi^-+p\rightarrow n+a$ depends on three axion couplings, $C_{ap}$, $C_{an}$, and the axion-pion-nucleon contact interaction $C_{a\pi N}$.  

Recently, the process $\pi^-+p\rightarrow n+a$ has been argued to be the dominating process to produce axions for a wide range of astrophysical conditions encountered inside supernovae~\cite{Carenza:2020cis,Fischer:2021jfm}.
The axion emissivity (the energy loss induced by axion emissions per unit volume and second) 
of this process
is given by
\bea
Q_a^{p\pi^-} &=&\int \prod_{\alpha=\pi, p, n, a}\frac{d^3 {\bf p}_\alpha}{(2\pi)^3 2 E_\alpha} 
\Big[ (2\pi)^4 \delta^{(4)}(p_{\pi}+p_{p} - p_n-p_a) \nonumber\\
&&\quad  \times\,  f_{\pi} (p_{\pi})  f_p(p_p) (1- f_n(p_n)) \, \sum_{s_p, s_n} |{\cal M}_{\pi^-+p\to n+a}|^2 \,  E_a \Big],
\eea
where  $p_\alpha = (E_\alpha , {\bf p}_\alpha)$ are the particle four-momenta, $f_\alpha(p_\alpha)$ are the Fermi-Dirac or Bose-Einstein distribution function, and $s_N$ ($N=p,n$) denotes the  nucleon spin.
Although the integrand has angular dependence after applying the energy-momentum conservation, 
 the nucleon distribution functions  can be approximated to be independent of those angles  in the non-relativistic limit.
Then, the squared matrix element can be first integrated over the relative angle of $\mathbf{p}_p$ with respect to $\mathbf{p}_{a}$, while the integration over the solid angle of $\mathbf{p}_a$ amounts to a factor of $4\pi$.\footnote{The squared matrix element depends on two independent angles, each from the Mandelstam variables $s$ and $u$.
We approximate these Mandelstam variables as  $s= (p_p+p_{\pi^-})^2 \simeq m_p^2 + m_{\pi^-}^2 + 2 m_p E_{\pi^-}$ and $u= (p_p-p_a)^2 \simeq m_p^2 - 2 m_p E_a$.}
%This is equivalent to keeping the leading and the next-to-leading order contributions 
%in the expansion in powers of  $1/m_N$.
%%e.g. the angles between $a$ and $p$, and $\pi^-$ and $p$.
%
%
%In the proton rest frame, we have
Taking the non-relativistic limit for the initial proton and integrating over the relative angle 
between $\mathbf{p}_{\pi^{-}}$ and $\mathbf{p}_a$, we find  
\bea 
\int d\Omega_{\pi^-} \sum_{s_p, s_n} |{\cal M}_{\pi^-+p\to n+ a}|^2 
=\frac{8\pi m_N^4}{f_a^2 f_\pi^2}  {\cal C}_a^{p\pi^-},
\eea 
where $m_N$ is the nucleon mass and 
${\cal C}_a^{p\pi^-}$ is a dimensionless quantity which can be expanded 
in powers of $1/m_N$ as
\bea \label{Ca2}
{\cal C}_a^{p\pi^-}&\simeq& \frac{2}{3}g_A^2 \left(\frac{|{\bf p}_{\pi}|}{m_N}\right)^2\left( 2C_+^2+C_{-}^2\right) 
+ \left(\frac{E_{\pi}}{m_N}\right)^2 C_{a\pi N}^2 
\nonumber\\
&& + \sqrt{2}g_A \left(\frac{E_{\pi}}{m_N}\right)^3 \left(1 -\frac{1}{3}\left( \frac{|{\bf p}_{\pi}|}{E_{\pi}}\right)^2 \right)   C_{a\pi N}C_{-},
\eea 
where  \bea
C_{\pm}=\frac{1}{2}\left(C_{ap}\pm C_{an}\right),  \quad E_{\pi} = \sqrt{m_{\pi^-}^2 + |{\bf p}_{\pi}|^2}.\eea
%Note that the second and third terms in the above expression of ${\cal C}_a^{p\pi^-}$, which represent %the effects of the axion-pion-nucleon contact interaction $C_{a\pi N}$,
%are always positive, so their contribution enhances the axion emission rate. 
We remark that the above expression of ${\cal C}_a^{p\pi^-}$ corresponds to the leading order result (in $1/m_N$) for which the three axion coupling  combinations, i.e., $2C_+^2+C^2_-$, $C_{a\pi N}C_-$ and $C_{a\pi N}^2$, are pretended to be independent parameters. 
%We also ignore medium effects, so treat all external states as free particles in the vacuum.
% and $C_{a\pi N}$ is given by Eq.~(\ref{Coeffs}).
%We calculate the matrix element in the limit that incoming and outgoing states are free particles.
As already noticed, $C_{a\pi N}$ is not an independent parameter, but is determined as
$C_{a\pi N}=\sqrt{2}C_-/g_A$ (see Eq.~(\ref{Coeffs})). 
Then the third term can be interpreted as a higher order term as it is suppressed compared to other terms
by additional power of $E_\pi/m_N$. 
However, our numerical estimation gives ${\cal C}_a^{p\pi^-} \simeq 0.02\, (2C_+^2+C_-^2) + 0.04 \,C_{a\pi N}^2+0.01 \,C_{a\pi N}C_-$ for typical parameter values, e.g., $T=40\,{\rm MeV}$, $|{\bf p}_{\pi}|\simeq \sqrt{3 m_{\pi^-} T} \simeq 130\,{\rm MeV}$. Note that the third term is comparable to the others. The relative importance of each term coincides with the final axion emissivity in Eq.~\eqref{fit:ppi} up to a small enhancement by the phase space integration.

%The third term originats from the interference between the third and the first two diagrams in %Fig.~\ref{diagram:pionproton} gets an extra suppression by $1/m_N$ compared to the others. 

For  $|\mathbf{p}_p|\gg |\mathbf{p}_{\pi, a}|$, 
the axion emissivity can be further approximated as  
\bea 
\hskip -1cm
Q_a^{p\pi^-} &\simeq &  \frac{ z_p z_{\pi^-} }{f_a^2 f_\pi^2}
\sqrt{\frac{m_N^7 T^{11}}{128\pi^{10}}}
 \int d x_p  \left( 
   \frac{x_p^2 e^{x_p^2}   }{ (e^{x_p^2} + z_n) (e^{x_p^2} + z_p)}\right) 
   \int dx_\pi  \left(\frac{x_\pi^2 \epsilon_\pi  {\cal C}_a^{p\pi^-} }{e^{\epsilon_\pi -y_\pi} - z_{\pi^-}} \right),
   \label{eq:Qapip}
\eea 
where $z_i= e^{(\mu_i-m_i)/T}$ are the fugacities, 
  $\epsilon_\pi = E_\pi/T$,  $y_\pi= m_{\pi^-}/T$, $x_\pi = |{\bf p}_{\pi}|/T$, and
 $x_p =|{\bf p}_p|/\sqrt{2m_N T}$.
 %Here $\gamma_{\rm sf}=\gamma_{\rm sf}(E_a)$  is the structure factor introduced to take into account %the multiple interactions of nucleons to change their spins, which turns out that
%its effect is not significant for $T^2\ll E_\pi^2\simeq E_a^2$ \cite{}.   
%
%Since ${\cal C}_a^{p\pi^-}$ contains the dependence on the temperature and the nucleon mass, the overall parametric dependence of $Q_a^{p\pi^-}$ would follow $\sim m_N^{1.5} T^{7.5}$ in Eq.~(\ref{eq:Qapip}).
%
%Inside the proto-neutron star, the calculation of the squared matrix amplitude in the massless pion %limit results in errors less than $10\%$
%$\mathcal{O}(10)\%$ errors 
%for 
%nucleon-nucleon bremsstrahlung \cite{Raffelt:1996wa, Carenza:2019pxu}, 
%the pion-nucleon scattering if we ignore the contact interactions.
% \cite{Fore:2019wib, Carenza:2020cis}.
%However, for the contribution from the contact interactions, there is an ${\cal O}(10)\%$ enhancement %compared to the case in the massless limit, so we should not neglect the pion mass effect. 
The emissivity Eq.~\eqref{eq:Qapip} depends on many astrophysical parameters which are related to each other by the equation of state, e.g., the temperature $T$ and the chemical potentials $\mu_i$ ($i=n,p,\pi^-$). To parameterize the astrophysical condition in terms of $T$ and the total mass density $\rho$, we use the fugacities obtained in \cite{Fore:2019wib} and numerically calculate the integral in
Eq.~(\ref{eq:Qapip}) around $T\sim 40$ MeV and $\rho\sim 10^{14}\,{\rm g}/{\rm cm}^3$. 
We then find
\begin{align}
\frac{Q_a^{p\pi^-}}{\text{erg}\cdot\text{cm}^{-3}\,\text{s}^{-1}}
&\simeq
1.4\times 10^{33} 
\,
T_{40}^{7.2}\rho_{14}^{1.1}
\left(\frac{10^9 \, {\rm GeV}}{f_a}\right)^2
\left(2C_+^2+C_{-}^2\right)
\nonumber
\\
&
\quad
+
2.3\times 10^{33}
\,
T_{40}^{6.6}\rho_{14}^{1.1}
\left(\frac{10^9 \, {\rm GeV}}{f_a}\right)^2
 C_{a\pi N}^2
\nonumber
\\
&
\quad
+
1.2\times 10^{33} 
\,
T_{40}^{7.5}\rho_{14}^{1.1}
\left(\frac{10^9 \, {\rm GeV}}{f_a}\right)^2
C_{a\pi N}C_-
,
\label{fit:ppi}
\end{align}
where $T_{40} \equiv T/(40\,{\rm MeV})$ and 
$\rho_{14} \equiv \rho/(10^{14}\,{\rm g}/{\rm cm}^3)$.
We stress that the above approximation is valid only for a narrow range of $T$ and $\rho$, i.e., for $T_{40}\in [0.9,1.1]$ and $\rho_{14}\in [1, 3]$, which is enough for our purpose to examine the effect of the axion-pion-nucleon contact interaction for an ambient condition inside supernovae.\footnote{The emissivity of $\pi^-+p\rightarrow n+a$ obtained in \cite{Carenza:2020cis} for $C_{a\pi N}=0$ is bigger than ours by a factor $\sim 2$. As the analysis of \cite{Carenza:2020cis} takes into account leading order medium effects, while ours does not, it is likely that this difference originates from medium effects.}
% We also recall
%that $C_{a\pi N}$ is not an independent variable, but determined by $C_{aN}$ ($N=n,p$) as $C_{a\pi N} %=  \sqrt{2}C_-/ g_A$ (see Eq.~(\ref{Coeffs})).

\begin{figure}[t!]
%\begin{center}
\includegraphics[width=0.496\textwidth]{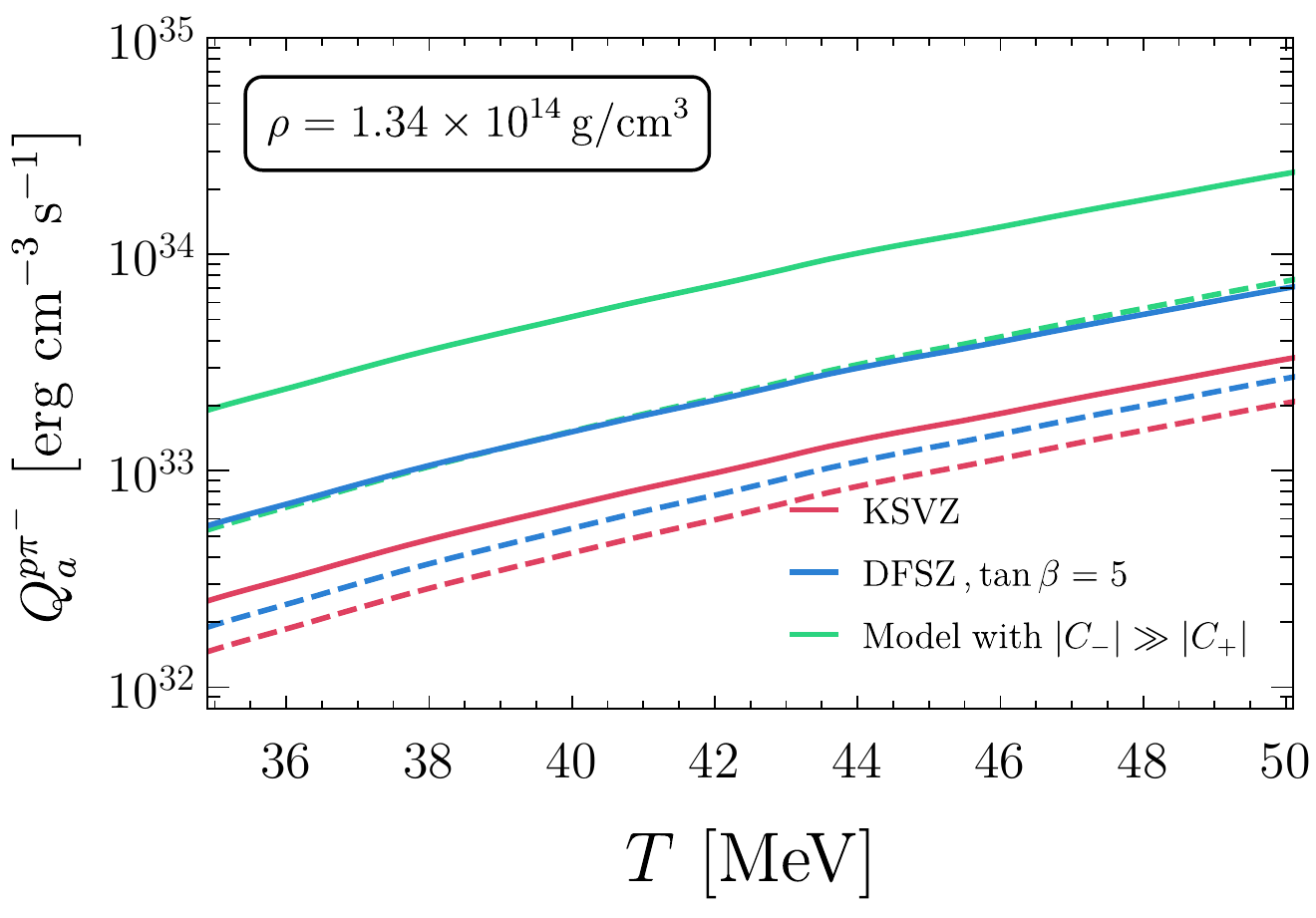} 
\includegraphics[width=0.496\textwidth]{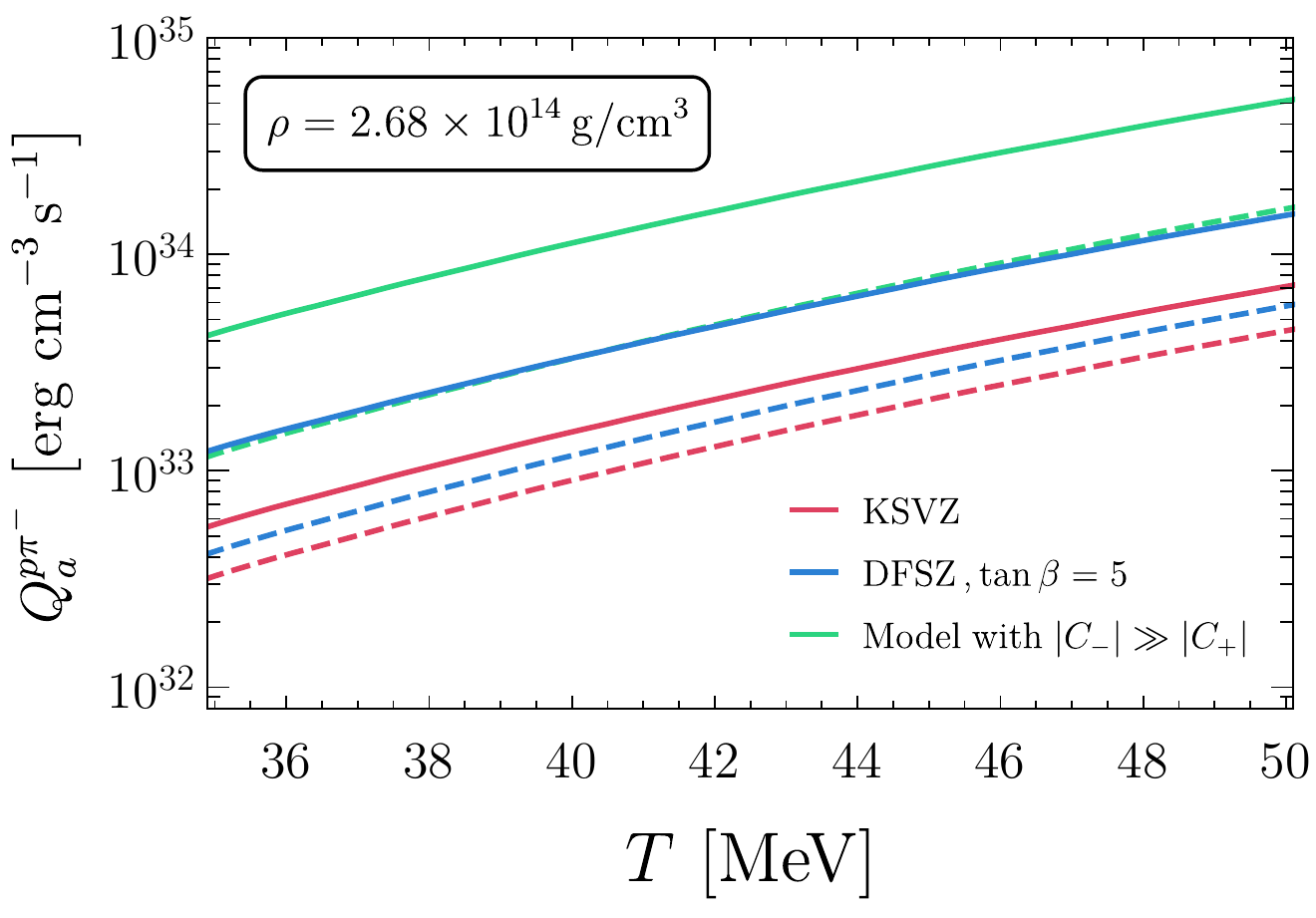} 
%\end{center}
\caption{Axion emissivities of $\pi^-+p\rightarrow n+a$  for the KSVZ, DFSZ, and a model 
with $|C_-|\gg |C_+|$.  All models are assumed to have $f_{a9}\equiv (f_a/c_G)/10^9\, {\rm GeV}=1$. The solid curves represent the total emissivity including the effect of the contact interaction $C_{a\pi N}$, while the dashed curves are  the emissivity without including the contribution from $C_{a \pi N}$.
}
% we include $c_G$ in the definition of $f_{a9}$ to normalize the axion-nucleon couplings for %comparison (see Eq.~(\ref{model:DFSZ})).
\label{fig:ppiCapiNratio}
\end{figure}

Obviously the second and third terms in the RHS of Eq.~\eqref{fit:ppi} represent the contributions to the axion emissivity from the axion-pion-nucleon contact interaction $C_{a\pi N}$. 
Because it is expected that the ratio is less sensitive to the uncertainty in nuclear physics,  we take the ratio between each term as 
\begin{align}
\frac{\Delta Q_{a,\,C_{a\pi N}^2}^{p\pi^-}}{\Delta Q_{a,\,(2C_+^2+C_{-}^2)}^{p\pi^-}} \simeq 2.0\left(\frac{C_-^2}{2C_+^2+ C_-^2}\right),
\quad
\frac{\Delta Q_{a,\,C_{a\pi N}C_-}^{p\pi^-}}{\Delta Q_{a,\,(2C_+^2+C_{-}^2)}^{p\pi^-}} \simeq 0.9\left( \frac{C_-^2}{2C_+^2+C_-^2}\right).
\end{align}
Here we use the relation $C_{a\pi N}=\sqrt{2}C_-/g_A$ in Eq.~\eqref{Coeffs}.  
 Therefore, the contributions from the contact interactions enhance the axion emissivity by $\mathcal{O}(1)$ in general.
We highlight in Fig.~\ref{fig:ppiCapiNratio} how much  $C_{a\pi N}$ enhances the axion emissivity for three benchmark axion models with $f_{a9}\equiv (f_a/c_G)/10^9\, {\rm GeV}=1$;
the KSVZ model of Eq.~\eqref{model:KSVZ} (red), the DFSZ model of Eq.~\eqref{model:DFSZ} with $\tan\beta=5$ (blue), and a model (green) to realize $|C_-|\gg |C_+|\simeq 0$ by satisfying the condition Eq.~\eqref{eq:maximalcase} for the PQ charges. For the third model, we choose 
$c_G=2$, $n_u=n_d=0$, $q_{H_u}=q_{H_d}=-1$, $\tan\beta=5$ for the model parameters in Eq.~\eqref{cucd},
which result in $C_{ap}\simeq -C_{an}\simeq -1.62$.\footnote{A simple way to realize such a case is to introduce PQ-charged exotic quarks in the minimal DFSZ model, which generate $\Delta c_G=-4$.}
Note that in our convention, the axion decay constant  $f_a$ is defined by the axion field range $a\cong a+2\pi f_a$, and the axion-gluon coupling is given by $c_G/f_a$ for an integer-valued parameter $c_G$.
We show that the contact interaction can enhance the axion emissivity by a factor $2-4$, depending on the pattern of axion couplings, and this conclusion will not change significantly when we include the corrections, e.g., the medium effects \cite{Carenza:2020cis,Fischer:2021jfm}. 
%Note thatthe ratio $Q_a^{p\pi^-} / (Q_a^{p\pi^-})_{C_{a\pi N}=0}$ has a mild temperature-dependence that arises mainly from the non-zero pion mass.

Since 
$z_p/z_n\sim z_{\pi^0}/z_{\pi^-}={\cal O}(0.1)$ inside proto-neutron star
\cite{Fore:2019wib}, the process $\pi^0+n\rightarrow n+a$ shown in Fig.~\ref{diagram:npi0na} can be as important as
$\pi^-+p\rightarrow n+a$.
Taking the same approach as  Eq.~(\ref{eq:Qapip}), 
the axion emissivity of $\pi^0 + n \to n + a$  can be approximated as 
%Fig.~\ref{diagram:npi0na}.
%
\bea 
Q_a^{n\pi^0 } &\simeq &  \frac{1}{2}\frac{ z_n z_{\pi^0} }{f_a^2 f_\pi^2}
\sqrt{\frac{m_N^7 T^{11}}{128\pi^{10}}}
 \int_0^\infty d x_{n} 
   \frac{x_n^2 e^{x_n^2}   }{ (e^{x_n^2} + z_n)^2}
   \int_0^\infty dx_\pi  \left(\frac{x_\pi^2 \epsilon_{\pi^0} \,  {\cal C}_a^{n\pi^0} }{e^{\epsilon_{\pi^0} -y_{\pi^0}} - z_{\pi^0}} \right),
   \label{eq:Qapin}
\eea 
where
$\epsilon_{\pi^0} = E_{\pi^0}/T$, $y_{\pi^0} = m_{\pi^0}/T$, and
\bea 
{\cal C}_a^{n\pi^0}&\simeq& \frac{4}{3}g_A^2 \left(\frac{|{\bf p}_{\pi}|}{m_N}\right)^2  C_{an}^2
.
\eea 
Using again the fugacities obtained in \cite{Fore:2019wib}, $Q_a^{n\pi^0 }$ can be further approximated as
\begin{align}
\frac{Q_a^{n\pi^0}}{\text{erg}\cdot\text{cm}^{-3}\,\text{s}^{-1}}
\simeq 
1.5\times 10^{33} 
\,
T_{40}^{7.5}\rho_{14}^{1.0}
\left(\frac{10^9 \, {\rm GeV}}{f_a}\right)^2
C_{an}^2
\label{fit:npi0}
\end{align}
for $T_{40}\in [0.9,1.1]$ and $\rho_{14}\in [1, 3]$. This shows that $Q_a^{n\pi^0}$
can be comparable to $Q_a^{p\pi^-}$ for $T\sim 40$ MeV and $\rho\sim 10^{14}\,{\rm g}/{\rm cm}^3$, {\it unless} $|C_{an}|\ll |C_{ap}|$.

It is also straightforward to confirm that the other pion-nucleon scattering processes, i.e.  $\pi^0+p\rightarrow p+a$ and $\pi^++n\rightarrow p+a$, give subleading contribution relative to
$\pi^-+p\rightarrow n+a$ and $\pi^0+n\rightarrow n+a$ for  $T\sim 40$ MeV and $\rho\sim 10^{14}\,{\rm g}/{\rm cm}^3$. For instance, for the process
$\pi^0+p\rightarrow p+a$, we find
\begin{align}
\frac{Q_a^{p\pi^0}}{\text{erg}\cdot\text{cm}^{-3}\,\text{s}^{-1}}
&\simeq  
2.7\times 10^{32} 
\,
T_{40}^{10.3}\rho_{14}^{0.56}
\left(\frac{10^9 \, {\rm GeV}}{f_a}\right)^2
C_{ap}^2
&\simeq \frac{z_p}{z_n}\frac{C_{ap}^2}{C_{an}^2} \frac{Q_a^{n\pi^0}}{\text{erg}\cdot\text{cm}^{-3}\,\text{s}^{-1}} 
,
\end{align}
where the fugacities of \cite{Fore:2019wib} are used for the last expression.
This shows that for $T_{40}\in [0.9,1.1]$ and $\rho_{14}\in [1, 3]$, $Q_a^{p\pi^0}<Q_a^{p\pi^-}$ over the entire axion parameter space.
%Checked: the ratio becomes minimal, 4.5, when $Can \simeq Cap/3$, and $T40 = 45/40$, $\rho14 = 1.34$.

\begin{comment}
\renewcommand{\arraystretch}{1.5}
\begin{table} 
\begin{center}
\begin{tabular}{ |c||c| }
\hline
 Models &  $Q_a^{n\pi^0} ~$\big[erg$\,\cdot\,$cm$^{-3}$\,s$^{-1}$\big]	
\\
 \hline\hline
KSVZ	&   	$3.0\times10^{30}\, T_{40}^{7.5} \rho_{14}^{1.0}f_{a9}^{-2}$	 	\\[0.5ex]
  \hline
%DFSZ 
%($\tan\beta=0.25$)	&   1.1 $T_{37}^{-0.026} \rho_{14}^{0.001}$	\\[0.5ex]
%  \hline
DFSZ 
($\tan\beta=1$)	&  	$5.4\times10^{30}\, T_{40}^{7.5} \rho_{14}^{1.0}f_{a9}^{-2}$	 	\\[0.5ex]
  \hline
%DFSZ 
%($\tan\beta=2$)	&   	&	 2.7  $T_{37}^{-0.32} \rho_{14}^{0.013}$	\\[0.5ex]
%  \hline
%DFSZ 
%($\tan\beta=10$)	&	&    2.9  $T_{37}^{-0.33} \rho_{14}^{0.013}$	\\[0.5ex]
%  \hline
DFSZ 
($\tan\beta=50$)	&	 $1.5\times10^{32}\, T_{40}^{7.5} \rho_{14}^{1.0}f_{a9}^{-2}$	 	\\[0.5ex]
%  \hline
%DFSZ 
%($\tan\beta=170$)	&   2.9 $ T_{37}^{-0.33} \rho_{14}^{0.013}  $	\\[0.5ex]
 \hline
$C_{ap}=-C_{an}=1/2$   	&	 $5.5\times10^{32}\, T_{40}^{7.5} \rho_{14}^{1.0}f_{a9}^{-2}$			 \\[0.5ex]
   \hline
\end{tabular}
\end{center} 
\caption{
The axion emissivity $Q^{n\pi^0}_a$ is estimated for four benchmark axion models. 
We define the parameters as
 $T_{40} \equiv T/(40\,{\rm MeV})$, 
$\rho_{14} \equiv \rho/(10^{14}{\rm g}/{\rm cm}^3)$,  $f_{a9}=(f_a/10^9\, {\rm GeV})$, and
$\tan\beta = \langle H_u\rangle/\langle H_d\rangle$.
As before, the fitting formulae are valid in the ranges of $T_{40}\in [0.9,1.1]$ and $\rho_{14}\in [1, 3]$.
}
\label{table:npi0}
\end{table}
\end{comment}

\subsection{Nucleon-nucleon bremsstrahlung}

\begin{figure}[t!]
\centering
\includegraphics[width=0.85\textwidth]{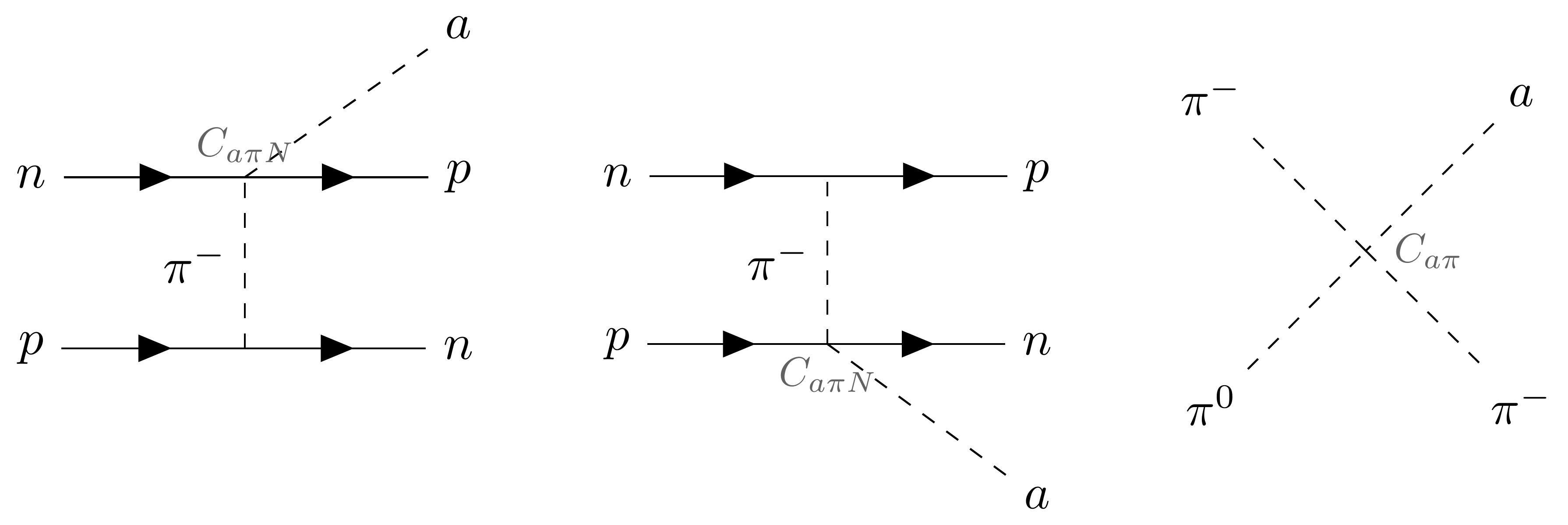} 
  \caption{Other diagrams involving the contact interactions.}
\label{diagram:others}
\end{figure}

For many years, the nucleon-nucleon bremsstrahlung has been considered to be the dominating process for axion emission  from supernovae.
Although a recent study indicates that the axion emissivity of the bremsstrahlung process is  sensitive to the corrections to the one-pion exchange as well as the medium effects~\cite{Carenza:2019pxu}, here we do a simpler analysis ignoring these corrections since we are mainly concerned with a relative importance of the axion-pion-nucleon contact interaction $C_{a\pi N}$ compared to the other axion-nucleon interactions.
In \cite{Carena:1988kr}, the same analysis has been done for the nucleon-nucleon bremsstrahlung with the contact interaction. It shows that the contribution from the contact interaction is negligible at the squared matrix element level. In this subsection, we examine the contribution from the contact interaction to the final axion emissivity including the phase space integration, and confirm that it is still negligible for the environmental parameters of SN~1987A.\footnote{In appendix~\ref{appendixA}, we estimate the axion emissivity in the degenerate limit by applying the analytic method presented in \cite{Iwamoto:1992jp}. In that estimation, the contact interaction seems to contribute to the axion emissivity in the same order of magnitude, but the environmental parameters given in \cite{Carenza:2020cis,Fore:2019wib} turn out to be not degenerate enough to apply the method.}

Among the three possible nucleon-nucleon bremsstrahlung processes, $n+n\rightarrow n+n+a$, $n+p\rightarrow n+p+a$, and $p+p\rightarrow p+p+a$,
at leading order in pion-nucleon couplings only the second process 
is affected by $C_{a\pi N}$ through the first two diagrams of Fig.~\ref{diagram:others}.
The axion emissivity of the three bremsstrahlung processes is given by 
\begin{align}
\label{eq:NNbremgeneral}
Q_a^I &= \int
\prod_{\substack{\alpha=N_1,N_2,\\N_3,N_4,\,a}}
\frac{d^3 {\bf p}_\alpha}{(2\pi)^3 2 E_\alpha} 
\Big[(2\pi)^4
\delta^{(4)}(p_1+p_2-p_3-p_4-p_a)
\nonumber\\
&
%\hspace{1.5cm}
\times
f_1(p_1) f_2(p_2)
(1-f_3(p_3))
(1-f_4(p_4))
\sum_\text{spins}	S_I	 |\mathcal{M}_I|^2
E_a
\Big]
\quad (I=nn, \, np, \, pp)
,
\end{align}
where 
$p_{1,2}=(E_{1,2}, \mathbf{p}_{1,2})$ and $p_{3,4}=(E_{3,4}, \mathbf{p}_{3,4})$ denote the initial and final nucleon four-momenta, $p_a=(E_a, \mathbf{p}_a)$ is the axion four-momentum, and 
$S_I$ is a symmetry factor for identical particles in the initial and final states, i.e., $S_{nn}=S_{pp}=1/4$ and $S_{np}=1$.
In the supernova environments, $|\mathbf{p}_a|\sim T\ll|\mathbf{p}_N| \sim \text{max}\left[\sqrt{m_N T}, \, p_F\right]$ where $p_F$ is the nucleon Fermi momentum.
Therefore we take the following approximation
\begin{align}
\label{eq:assumptionNN}
\mathbf{p}_1+\mathbf{p}_2 \simeq \mathbf{p}_3 + \mathbf{p}_4\,,
\end{align}
which simplifies the kinematics significantly.
With this approximation and also at leading order in $1/m_N$, the squared matrix elements averaged over the axion momentum direction are given by 
\begin{align}
\left\langle\sum_\text{spins}
|\mathcal{M}_{nn}|^2\right\rangle
&=
\frac{16}{3}
\frac{g_A^4}{f_\pi^4}
\frac{m_N^4}{f_a^2}
C_{an}^2
\left(
\frac{|\mathbf{k}|^4}{(|\mathbf{k}|^2+m_\pi^2)^2}
+
\frac{|\mathbf{l}|^4}{(|\mathbf{l}|^2+m_\pi^2)^2}
\right.
\nonumber\\
&
\left.
\hspace{3.5cm}
+
(1-\beta) \frac{|\mathbf{k}|^2 |\mathbf{l}|^2}{(|\mathbf{k}|^2+m_\pi^2)(|\mathbf{l}|^2+m_\pi^2)}\right)
,
\\
\left\langle\sum_\text{spins}
|\mathcal{M}_{pp}|^2
\right\rangle
&=
\frac{16}{3}
\frac{g_A^4}{f_\pi^4}
\frac{m_N^4}{f_a^2}
C_{ap}^2
\left(
\frac{|\mathbf{k}|^4}{(|\mathbf{k}|^2+m_\pi^2)^2}
+
\frac{|\mathbf{l}|^4}{(|\mathbf{l}|^2+m_\pi^2)^2}
\right.
\nonumber\\
&
\left.
\hspace{3.5cm}
+
(1-\beta) \frac{|\mathbf{k}|^2 |\mathbf{l}|^2}{(|\mathbf{k}|^2+m_\pi^2)(|\mathbf{l}|^2+m_\pi^2)}\right)
,
\\
\left\langle\sum_\text{spins}|\mathcal{M}_{np}|^2
\right\rangle
&=
\frac{16}{3}
\frac{g_A^2}{f_\pi^4}
\frac{m_N^4}{f_a^2}
\left[
g_A^2 
\left\{
(4C_+^2+2C_-^2) \frac{|\mathbf{k}|^4}{(|\mathbf{k}|^2+m_\pi^2)^2}
+
 (C_+^2+C_-^2) \frac{|\mathbf{l}|^4}{(|\mathbf{l}|^2+m_\pi^2)^2}
 \right.
\right.
\nonumber
\\
&
\left.
\hspace{2.2cm}
- 2 \left( (C_+^2+C_-^2)- (3C_+^2+C_-^2)\,\frac{\beta}{3}\right)
\frac{|\mathbf{k}|^2 |\mathbf{l}|^2}{(|\mathbf{k}|^2+m_\pi^2)(|\mathbf{l}|^2+m_\pi^2)}
\right\}
\nonumber
\\
&
\hspace{2.5cm}
%+ 36 g_A^2 C_-^2 \frac{\omega_a^2|\mathbf{k}|^2}{(|\mathbf{k}|^2+m_\pi^2)^2}
%\nonumber
%\\
%&
%\hspace{1.7cm}
%\left.
%+
%36\sqrt{2}g_A C_{a\pi N}C_- \frac{\omega_a^2|\mathbf{k}|^2}{(|\mathbf{k}|^2+m_\pi^2)^2}
\left.
+
3 C_{a\pi N}^2 \frac{|\mathbf{p}_a|^2|\mathbf{k}|^2}{(|\mathbf{k}|^2+m_\pi^2)^2}
\right],
\label{np:matrixelem}
\end{align}
where 
\begin{align}
\mathbf{k} \equiv \mathbf{p}_1 - \mathbf{p}_3,
\quad
\mathbf{l} \equiv \mathbf{p}_1 - \mathbf{p}_4,
\quad
\beta \equiv 3\left(\frac{{\mathbf{k}}\cdot{\mathbf{l}}}{\left|\mathbf{k}\right|\left|\mathbf{l}\right|}\right)^2.
%\quad
%C_\pm \equiv \frac{C_{ap}\pm C_{an}}{2}
\end{align}
For the neutron-proton bremsstrahlung, we define the momentum exchanges as $\mathbf{k} = \mathbf{p}_n^i - \mathbf{p}_p^f$ and $\mathbf{l} = \mathbf{p}_n^i -  \mathbf{p}_n^f$.
%Note that the definition $C_-$ has the opposite sign compared to \cite{Carenza:2019pxu}. It is due to make it have the same sign with $C_{a\pi N}$ (see Eq.~(\ref{Coeffs})).
%
%

While the squared matrix elements of $n+n\rightarrow n+n+a$ and $p+p\rightarrow p+p+a$ are the same as  the previous results~\cite{Carenza:2019pxu}, the squared matrix element of $n+p\rightarrow n+p+a$ includes
% i.e., Eq.~\eqref{np:matrixelem}, 
an additional contribution from the contact interaction $C_{a\pi N}$.
We remark that we have only displayed the leading-order contribution (in $1/m_N$) for each coupling term in the angle-averaged squared matrix elements.
% terms in the $|\mathbf{p}_a|/|\mathbf{p}_N|\ll1$ limit for each of the contributions from the %couplings $C_\pm$ and $C_{a\pi N}$
%in Eq.~\eqref{np:matrixelem}. 
Then, compared to other terms,  the term induced by $C_{a\pi N}$ in Eq.~\eqref{np:matrixelem} is intrinsically higher order as it is  suppressed by  $|\mathbf{p}_a|^2/ 
|\mathbf{p}_N|^2\sim T/m_N$ for $|\mathbf{k}|\sim |\mathbf{l}|\sim |\mathbf{p}_N|$. 
%Meanwhile, we comment that there may exists contributions from the general coupling combinations, %e.g., $C_\pm^2$, $C_+C_-$, and $C_\pm C_{a\pi N}$, that are suppressed by the same factor.\footnote{At %the level of squared matrix element before averaging over the axion momentum direction, there exist %contributions that are less suppressed, i.e., by a factor of $|\mathbf{p}_a|/|\mathbf{p}_N|$. However, %such terms have vanishing contribution to the axion emissivity since they depend linearly on the axion %angles $\cos\theta_{ai}$.}
This indicates that  the contribution from $C_{a\pi N}$ to the axion emissivity of $n+p\rightarrow n+p+a$ is likely to be negligible as pointed out in \cite{Carena:1988kr}.
If the typical values into the kinetic parameters are taken, e.g., $T= 40\,{\rm MeV}$, $\beta \simeq 1.3$ (non-degenerate limit), $|\mathbf{k}|\sim |\mathbf{l}|\sim |\mathbf{p}_N|\simeq \sqrt{3 m_N T}\simeq 340\,{\rm MeV}$, $|\mathbf{p}_a|\simeq E_a \simeq |\mathbf{p}_N|^2/(2m_N) \simeq 60\,{\rm MeV}$, we could see a numerical estimate of the square brackets in Eq.~\eqref{np:matrixelem}, $\left[\,\cdots \right]\simeq 6.6 \,C_+^2 + 2.2\,C_-^2+0.07 \,C_{a\pi N}^2$. The estimation predicts a relative importance of each term, which is shown in Eq.~\eqref{fit:np}, although there appears some enhancement of the contribution from the contact interaction after the phase space integration.

%Such contributions are sub-leading compared to the leading-order contributions from the $C_\pm$ %coupling, but they are comparable to the leading-order contribution from the $C_{a\pi N}$ coupling.
%Nevertheless, we confirm that any terms that are suppressed by the factor of $|\mathbf{p}_a|^2/|
%\mathbf{p}_N|^2$ give negligible contributions to the axion emissivity in supernova environments;
%they are smaller by the factor of $\sim T/m_N$ as expected. %compared to the leading order contributions from the $C_\pm$ coupling.

%Since we take the approximation neglecting the axion momentum, i.e., $|\mathbf{k}_a|/|\mathbf{p}_N|$, there could be a higher order contribution for each term. In Eq.~(\ref{np:matrixelem}), we keep the leading order of them.

The axion emissivity in Eq.~(\ref{eq:NNbremgeneral}) can be simplified 
by taking non-relativistic limit for nucleons together with the approximation Eq.~(\ref{eq:assumptionNN}).
% and by taking the non-relativistic limit, i.e., $m_N \gg |\mathbf{p}_N|$. 
%A proper choice of dimensionless variables in the center-of-mass frame~
Following \cite{Brinkmann:1988vi,Raffelt:1993ix},
we can write the axion emissivities in a form which allows a numerical calculation of the phase space integration:
%For $I=nn, \, np, \, pp$,
\begin{align}
Q_a^I &\simeq \sqrt{\frac{m_NT^{13}}{2^{9}\pi^{16}}}
\int_0^\infty
du_+
\int_0^\infty
du_-
\int_{-1}^1
d\gamma_{+-}
\int_0^{u_-}
d u_{3c}
\int_{4\pi}
d\Omega_{3c}
\sqrt{u_+u_- u_{3c}}
(u_- - u_{3c})^2
\nonumber\\
&
\hspace{1.5cm}
\times
f_1 f_2
(1-f_3)
(1-f_4)
\,
\sum_\text{spins}	S_I \left\langle |\mathcal{M}_I|^2 \right\rangle_{_{\mathbf{p}_{4c}=-\mathbf{p}_{3c},\, E_a = 2T(u_- - u_{3c})}}
%\frac{1}{4\pi}
%\int_{4\pi} d \Omega_a
%\sum_\text{spins}	S_I	 |\mathcal{M}_I|^2
%\bigg|_{\mathbf{p}_{4c}=-\mathbf{p}_{3c},\, \omega_a = 2T(u_- - u_{3c})}
\,,
\label{eq:QaInumerics}
\end{align}
where 
\begin{align}
u_i \equiv \frac{\mathbf{p}_i^2}{2m_N T},
\quad
\mathbf{p}_{\pm} \equiv \frac{\mathbf{p}_1 \pm \mathbf{p}_2}{2},
\quad
\mathbf{p}_{j c} \equiv \mathbf{p}_j- \mathbf{p}_+,
\quad
\gamma_{kl} \equiv \frac{\mathbf{p}_k\cdot \mathbf{p}_l}{|\mathbf{p}_k| |\mathbf{p}_l|}.
\end{align}
Again, we use the fugacities of nucleons from \cite{Fore:2019wib} to numerically calculate the above axion emissivities, which results in\footnote{Our numerical results agree well with the analytic results of the previous works~\cite{Brinkmann:1988vi,Iwamoto:1992jp}; for the contributions from $C_\pm$, the agreement is at the level of ${\cal O}(10)\%$ discrepancy in both degenerate ($z_{n/p}\gg1$) and non-degenerate ($z_{n/p}\ll1$) limits.
We also confirm that the contribution from $C_{a\pi N}$ agrees well with an analytic result in the degenerate limit \cite{Iwamoto:1992jp}. See appendix~\ref{appendixA}.}
%Given the information of the chemical potentials in the supernova environments~\cite{Fore:2019wib}, 
%we apply Eq.~(\ref{eq:QaInumerics}) for each process and find the fitting functions with respect to %the temperature and the density of the supernova in the relevant range, i.e., 
%
\begin{align}
\frac{Q_a^{nn}}{\text{erg}\cdot\text{cm}^{-3}\,\text{s}^{-1}}
&\simeq
3.5\times 10^{33}\,
T_{40}^{3.9}\rho_{14}^{2.1}
\left(\frac{10^9 \,{\rm GeV}}{f_a}\right)^2
C_{an}^2
,
\label{fit:nn}
\\
\frac{Q_a^{np}}{\text{erg}\cdot\text{cm}^{-3}\,\text{s}^{-1}}
&\simeq
7.5\times 10^{33}\,
T_{40}^{6.9}\rho_{14}^{1.5}
\left(\frac{10^9 \,{\rm GeV}}{f_a}\right)^2
C_+^2
\nonumber\\
&
\hspace{0.5cm}
+
2.5\times 10^{33}\,
T_{40}^{6.9}\rho_{14}^{1.5}
\left(\frac{10^9 \,{\rm GeV}}{f_a}\right)^2
C_-^2
\label{fit:np}
\\
&\hspace{0.5cm}
+
2.7\times 10^{32} \,
T_{40}^{7.9}\rho_{14}^{1.5}
\left(\frac{10^9 \,{\rm GeV}}{f_a}\right)^2
C_{a\pi N}^2
,
\nonumber
\\
\frac{Q_a^{pp}}{\text{erg}\cdot\text{cm}^{-3}\,\text{s}^{-1}}
&\simeq
9.9\times 10^{31}\,
T_{40}^{9.9}\rho_{14}^{0.92}
\left(\frac{10^9 \,{\rm GeV}}{f_a}\right)^2
C_{ap}^2\,
\label{fit:pp}
\end{align}
%where
%$C_\pm \equiv (C_{ap}\pm C_{an})/2$.
%
for $T_{40}\in [0.9,1.1]$ and $\rho_{14}\in [1, 3]$.
%Since $C_{a\pi N} =  (C_{ap} - C_{an})/(\sqrt{2} g_A) = \sqrt{2}C_-/g_A$ (see Eq.~(\ref{Coeffs})),

\begin{figure}[t!]
%\begin{center}
\includegraphics[width=0.496\textwidth]{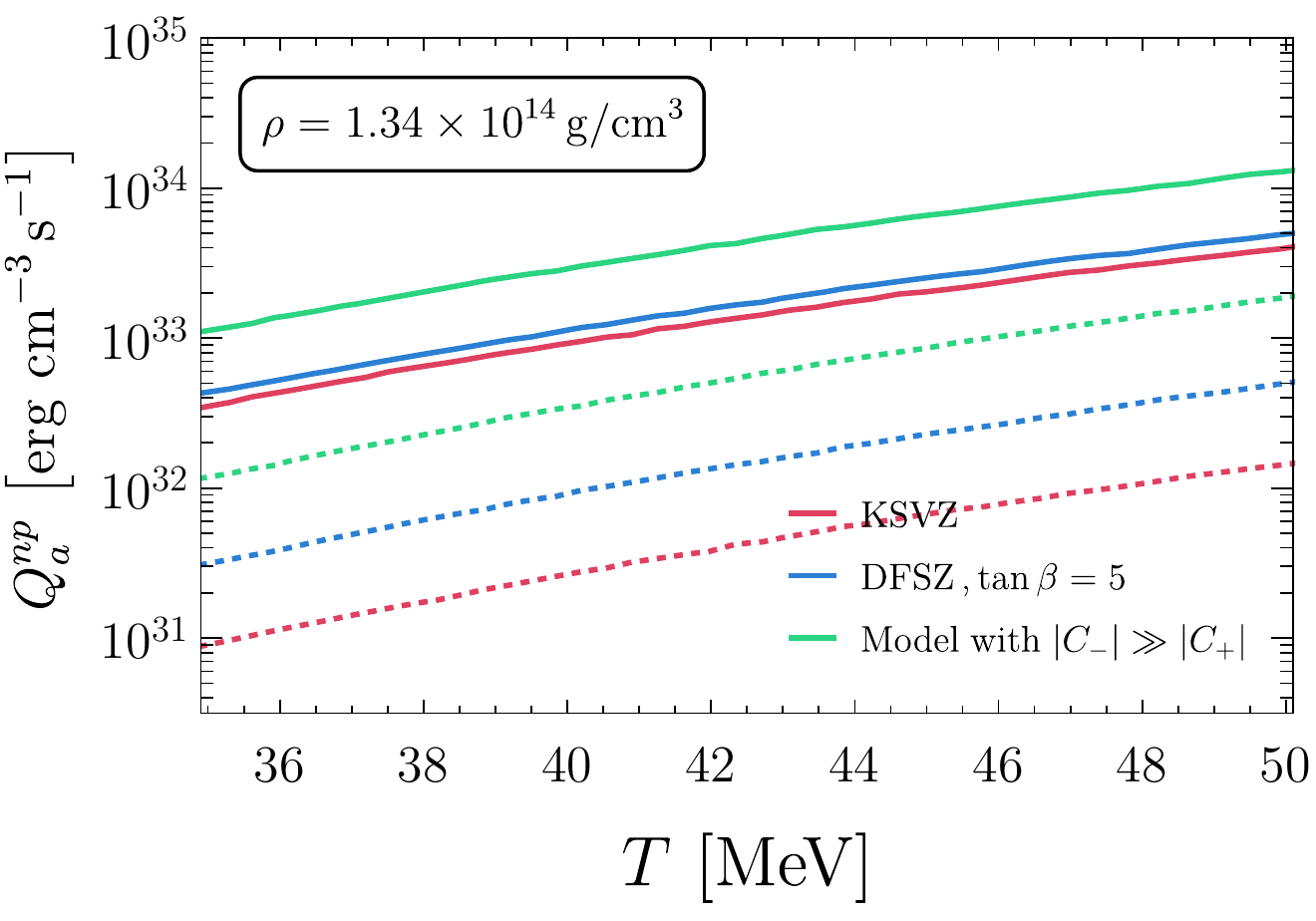} 
\includegraphics[width=0.496\textwidth]{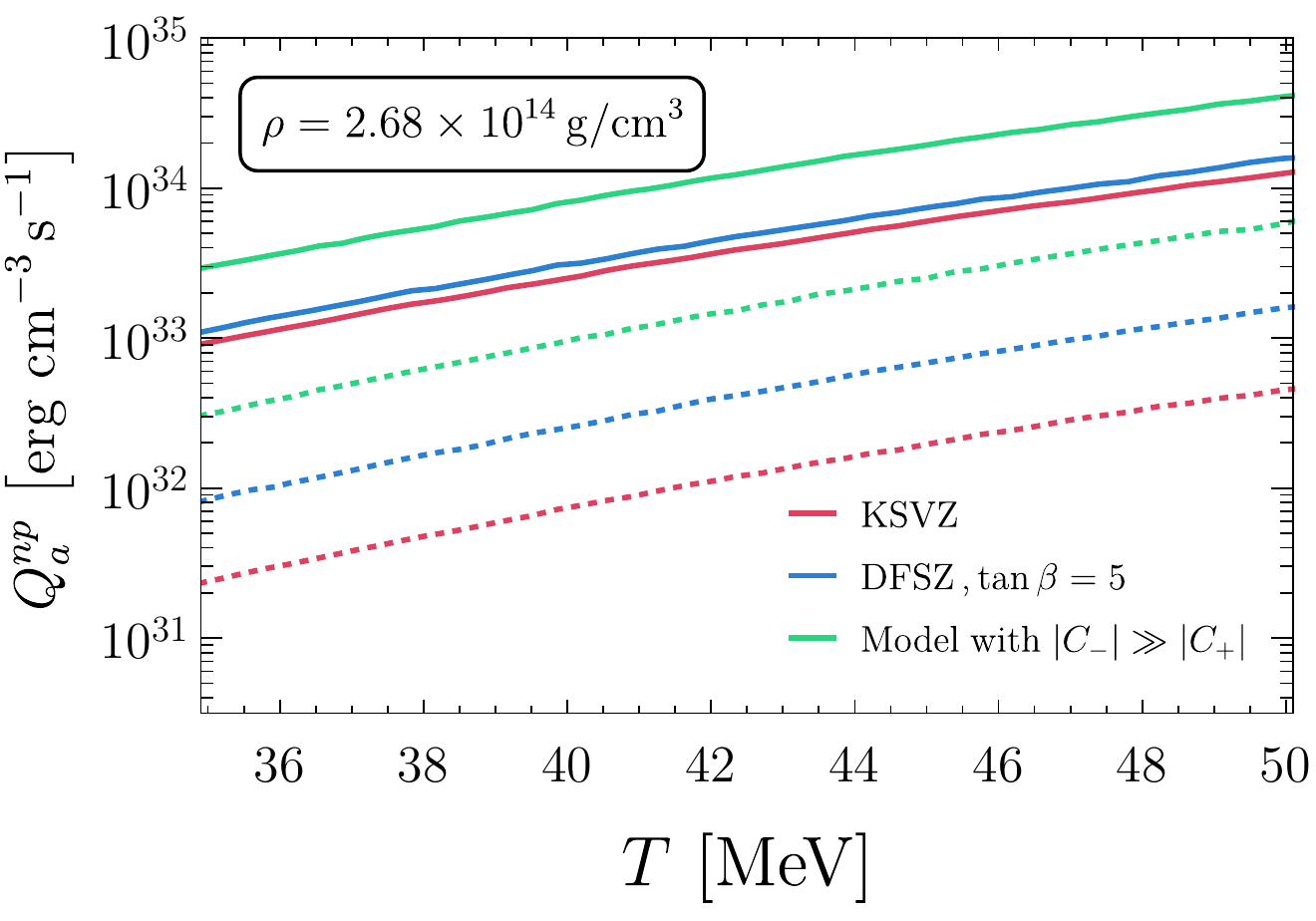} 
%\end{center}
\caption{The axion emissivity $Q^{np}_a$  for the three benchmark models considered in Fig.~\ref{fig:ppiCapiNratio}. 
The solid (dotted) curves correspond to the total value of $Q^{np}_a$ (the piece induced only by $C_{a\pi N}$).
}
\label{fig:npCapiNratio}
\end{figure}

The above result shows that, as anticipated from the structure of the squared matrix element, the contribution to $Q^{np}_a$ from $C_{a\pi N}$ is indeed about one order of magnitude smaller than the contribution from $C_-$ 
for astrophysical environments with $T\sim 40$ MeV and $\rho\sim 10^{14} \, {\rm g}/{\rm cm}^3$.
Using the relation $C_{a\pi N}=\sqrt{2}C_-/g_A$, the ratios between the contribution from the contact interaction and the other terms become, respectively,
\begin{align}
\frac{\Delta Q_{a,\,C_{a\pi N}^2}^{np}}{\Delta Q_{a,\,C_+^2}^{np}} \simeq 0.04 \left(\frac{C_{a\pi N}^2}{C_+^2}\right),\quad
\frac{\Delta Q_{a,\,C_{a\pi N}^2}^{np}}{\Delta Q_{a,\,C_{-}^2}^{np}} \simeq 0.1.
\end{align}
 As in the case of $Q_a^{p\pi^-}$, these ratios are expected to be less sensitive to the corrections beyond the one-pion exchange and the medium effects, so the effect of the contact interaction $C_{a\pi N}$ on the nucleon-nucleon bremsstrahlung is negligible.
%We find that the contact interaction for the neutron-proton bremsstrahlung always gives a subdominant %contribution for the environment conditions that we consider, i.e., $T_{40}\in [0.9,1.1]$ and $%\rho_{14}\in [1, 3]$;
In Fig.~\ref{fig:npCapiNratio}, we compare the total value of the axion emissivity $Q^{np}_a$ (solid curves) with the piece $\Delta Q_{a,\,C_{a\pi N}^2}^{np}$ (dotted curves) induced only by $C_{a\pi N}$ for the three benchmark models considered 
in Fig.~\ref{fig:ppiCapiNratio}.
The result shows that the contribution from $C_{a\pi N}$ is negligible for $30\lesssim T/{\rm MeV} \lesssim 50$ and $1\lesssim \rho/(10^{14} \,{\rm g}/{\rm cm}^3)\lesssim 3$, which is expected to be true for even wider range of $T$ and $\rho$. 
Note that $Q^{np}_a$ is comparable to (or even larger than) $Q^{nn}_a$, although the proton number density is significantly smaller than the neutron number density.
This is partly due to the symmetry factor $S$ compensating the small proton fraction.\footnote{The relative importance of the neutron-proton bremsstrahlung is discussed within the framework of the neutrino emission through the nucleon-nucleon bremsstrahlung~\cite{Yakovlev:2000jp}.}
%We finally remark that although the axion emissivity of the nucleon-nucleon bremsstrahlung processes are known to be sensitive to medium effects \cite{Carenza:2019pxu}, it is likely that the ratio $\Delta Q^{np}_a/Q^{np}_a$ is   less sensitive. We therefore expect that the effect of the contact interaction $C_{a\pi N}$ on the axion emissivity of the nucleon-nucleon bremsstrahlung processes remains to be negligible even when medium effects are included  as in \cite{Carenza:2020cis,Fischer:2021jfm}.

%

%

\subsection{Pion-pion scattering: $\pi+\pi\rightarrow \pi+a$}

Let us finally consider the possible consequence of the axion-pion contact interaction  $C_{a\pi}$ in Eq.~(\ref{Lhadron}). Axions can be produced by this coupling through the pion-pion scattering process $\pi^-+\pi^0\to \pi^- + a$ (see the third diagram in Fig.~\ref{diagram:others}).
%
%There also exists a process $N+\pi \rightarrow N+\pi + a$ through the $C_{a\pi}$ coupling, but it should be subdominant to the neutron-proton bremsstrahlung because the number density of $\pi^-$ is one order of magnitude smaller than that of the proton \cite{Fore:2019wib}, and the couplings have the same structure, $C_{a\pi N} \sim C_{a\pi}\sim (C_{ap}-C_{an})/2$ (see Eq.~(\ref{Coeffs})).
%we do not expect any particular structure of models which enhances only $C_{a\pi}$ coupling.
%
The corresponding  emissivity can be simplified without any kinematic approximation as follows:
\begin{align}
Q_a^{\pi^-\pi^0} 
=&
\frac{9}{2^{12}\pi^7 } \frac{C_{a\pi}^2}{f_\pi^2f_a^2}z_{\pi^-} z_{\pi^0} T^9\int dx_\text{in} dx_0 d\Omega_{\pi^-}^\text{in} d\Omega_{\pi^0}
\frac{x_\text{in}^2}{\sqrt{y_\pi^2+x_\text{in}^2}}\frac{x_0^2}{\sqrt{y_\pi^2+x_0^2}}\frac{x_a^2}{E_{\rm out}/T}
\left(\frac{\mathbf{p}_{\pi^0}\cdot \mathbf{p}_a}{T^2}\right)^2
\nonumber\\
&\hspace{1.5cm}\times
\frac{1}{e^{\sqrt{y_\pi^2+x_\text{in}^2}-y_\pi}-z_{\pi^-}}\frac{1}{e^{\sqrt{y_\pi^2+x_0^2}-y_\pi}-z_{\pi^0}}
\frac{e^{E_{\rm out}/T-y_\pi}}{e^{E_{\rm out}/T-y_\pi}-z_{\pi^-}}
,
\label{eq:Qpionpionprocess}
\end{align}
where
$x_{\rm in} = {|\mathbf{p}_{\pi^-}^\text{in}|}/{T}$ for the incoming $\pi^-$,
$x_0 = {|\mathbf{p}_{\pi^0}|}/{T}$,
$x_a = {|\mathbf{p_a}|}/{T}$,
$y_\pi = {m_\pi}/{T}$, and finally
$E_{\rm out}
=
\sqrt{m_\pi^2+(\mathbf{p}_{\pi^-}^{\text{in}}+\mathbf{p}_{\pi^0}-\mathbf{p}_a)^2}$ is the energy of the outgoing $\pi^-$.
%In the expression, the subscript or superscript ``in" and the subscript ``out" denote the negatively %charged pions in the initial and the final states, respectively. Here, the masses of neutral and %charged pions are approximated to be equal.
%
Like the emissivity of other processes, we use the fugacities of pions from \cite{Fore:2019wib} and calculate the integral in Eq.~(\ref{eq:Qpionpionprocess}) numerically to find
\begin{align}
\frac{Q_a^{\pi^-\pi^0}}{\text{erg}\cdot\text{cm}^{-3}\,\text{s}^{-1}}
&
=
3.6\times 10^{31}\,
T_{40}^{10.1}\rho_{14}^{1.1}
\left(\frac{10^9\, {\rm GeV}}{f_a}\right)^2
C_{a\pi}^2.
\label{fit:pipi}
\end{align}
%
%
%By applying parameter values for $C_{a\pi}$ in each model,
% \begin{align}
% Q_a^{\pi^-\pi^0}\Big|_{\rm KSVZ} &= 1.5\times 10^{30} \, {\rm erg}\,{\rm cm}^{-3}  {\rm s}^{-1},
% \nonumber\\
%  Q_a^{\pi^-\pi^0}\Big|_{\rm DFSZ} 
%  &= 
%  1.5\times 10^{30} \, {\rm erg}\,{\rm cm}^{-3}  {\rm s}^{-1} \qquad (\tan\beta = 1),
%\nonumber  \\
%&= 5.8\times 10^{30} \, {\rm erg}\,{\rm cm}^{-3}  {\rm s}^{-1}		\qquad (\tan\beta = 50)
%,
%\\
%  Q_a^{\pi^-\pi^0}\Big|_{C_{ap}=-C_{an}=1/2} 
%&=
%7.6\times 10^{30} \, {\rm erg}\,{\rm cm}^{-3}  {\rm s}^{-1}		
%  .
% \nonumber
 %\end{align}
%
The above result shows that the axion emissivity of the pion-pion scattering $\pi^-+\pi^0\to \pi^- + a$ 
is negligible compared to that of $\pi^-+p\to n+a$ for $T\sim 40$ MeV and $\rho\sim 10^{14}\,{\rm g}/{\rm cm}^3$ and $C_{a\pi} = 2(C_{ap} -C_{an})/(3 g_A)$ (see Eq.~(\ref{Coeffs})).
%Accordingly, the pion scattering process is negligible for the axion emission in general.

\section{Conclusions and Discussion\label{Sec:Conc}}

In this paper, we have studied the axion emission from supernovae with a complete set of relevant axion couplings including the axion-pion-nucleon contact interaction $C_{a\pi N}$ and the axion-pion contact interaction $C_{a\pi}$ in Eq.~(\ref{Lhadron}). 
A recent study suggests that the abundance of negatively charged pions inside supernovae is significantly enhanced by the strong interactions \cite{Fore:2019wib}, indicating that
the  pion-induced process $\pi^-+p\rightarrow a+n$ is the dominating  process for a wide range of astrophysical conditions encountered inside supernovae \cite{Carenza:2020cis,Fischer:2021jfm}.  We thus examined how this pion-induced process is affected by $C_{a\pi N}$. We also examined the effect of $C_{a\pi N}$ on
the nucleon-nucleon bremsstrahlung which has been considered as the dominating process for many years.

Since we are mainly concerned with the role of the two previously ignored couplings $C_{a\pi N}$ and $C_{a\pi}$, we have focused on the axion coupling dependence of the axion emissivity within a simple approximation to keep only the leading order in pion-nucleon couplings, which also ignores medium effects. In such an approximation, we could show the axion coupling dependence more explicitly and examine the ignored couplings for three processes. Two processes, $\pi^-+p\rightarrow a+n$ 
and $n+p\to n+p+a$, are affected by $C_{a\pi N}$, and the pion-pion scattering, $\pi+\pi\rightarrow \pi+a$, is affected by $C_{a\pi}$.  We found that $C_{a\pi N}$
can enhance the
axion emissivity of $\pi^-+p\rightarrow a+n$ by a factor of $2-4$, depending  
on the pattern of axion couplings determined by the underlying axion model, while there is no
substantial effect on $n+p\rightarrow n+p+a$. Although it is independent of $C_{a\pi N}$, we have also examined the axion emissivity of $\pi^0+n\to n+a$ and find that it can be comparable to the emissivity of $\pi^-+p\rightarrow a+n$ over a wide range of axion parameter space.
For the axion-pion contact interaction $C_{a\pi}$, we
find that the corresponding axion emissivity is always negligible compared to that of $\pi^-+p\rightarrow a+n$ for ambient conditions encountered inside supernovae.

Let us make final remarks on the approximation we made. For the matrix elements, the higher-order diagrams could give comparable contributions due to the strong interaction. Moreover, the medium effects significantly change the axion emissivity, particularly for the nucleon-nucleon bremsstrahlung~\cite{Carenza:2019pxu, Fischer:2021jfm}. 
However, even including these effects,
the relative contribution of the axion-pion-nucleon contact interaction $C_{a\pi N}$ to the axion emissivity would remain similar because 
it is likely that the ratio $\Delta Q^I_{a,\,C_{a\pi N}}/Q^I_a$ is less sensitive to the corrections than the emissivity itself, where $Q^I_a$ is the axion emissivity of the $I$-th process, and $\Delta Q^I_{a,\,C_{a\pi N}}$ is the part of $Q^I_a$ induced by $C_{a\pi N}$.
%Thus, our results should be interpreted with some caution. 
One of the purposes of this work is to call attention to the possible importance of the contact interactions which have been neglected so far. 
%We have derive the enhancement factor for some popular axions models including
%the KSVZ and DFSZ models,
%In this paper, these effects are neglected since,
% they would not change the order of magnitude, and
%  \cite{Carenza:2020cis, 2108.13726}
%It has been noticed that those additional effects can significantly affect the axion emissivity of each processes. 
It would be interesting to perform the analysis taking into account more precise matrix elements and medium effects with a complete set of axion couplings.
We will investigate this issue in a self-consistent way for both the pion-nucleon scattering and the nucleon-nucleon bremsstrahlung in future works.
%Our analysis may also have implication for other stellar objects such as neutron star merger.
%, as well as its model dependence.
%

\section*{Acknowledgments}
 
This work was supported by IBS under the project code, IBS-R018-D1.
We are grateful to S. Yun for helpful discussions, especially on the nucleon-nucleon bremsstrahlung process in the early stage of this project.

\appendix

\section{Axion emissivity from $n+p\rightarrow n+p+a$ in degenerate limit \label{appendixA}}

\begin{figure}[t]
\begin{center}
\includegraphics[width=0.65\textwidth]{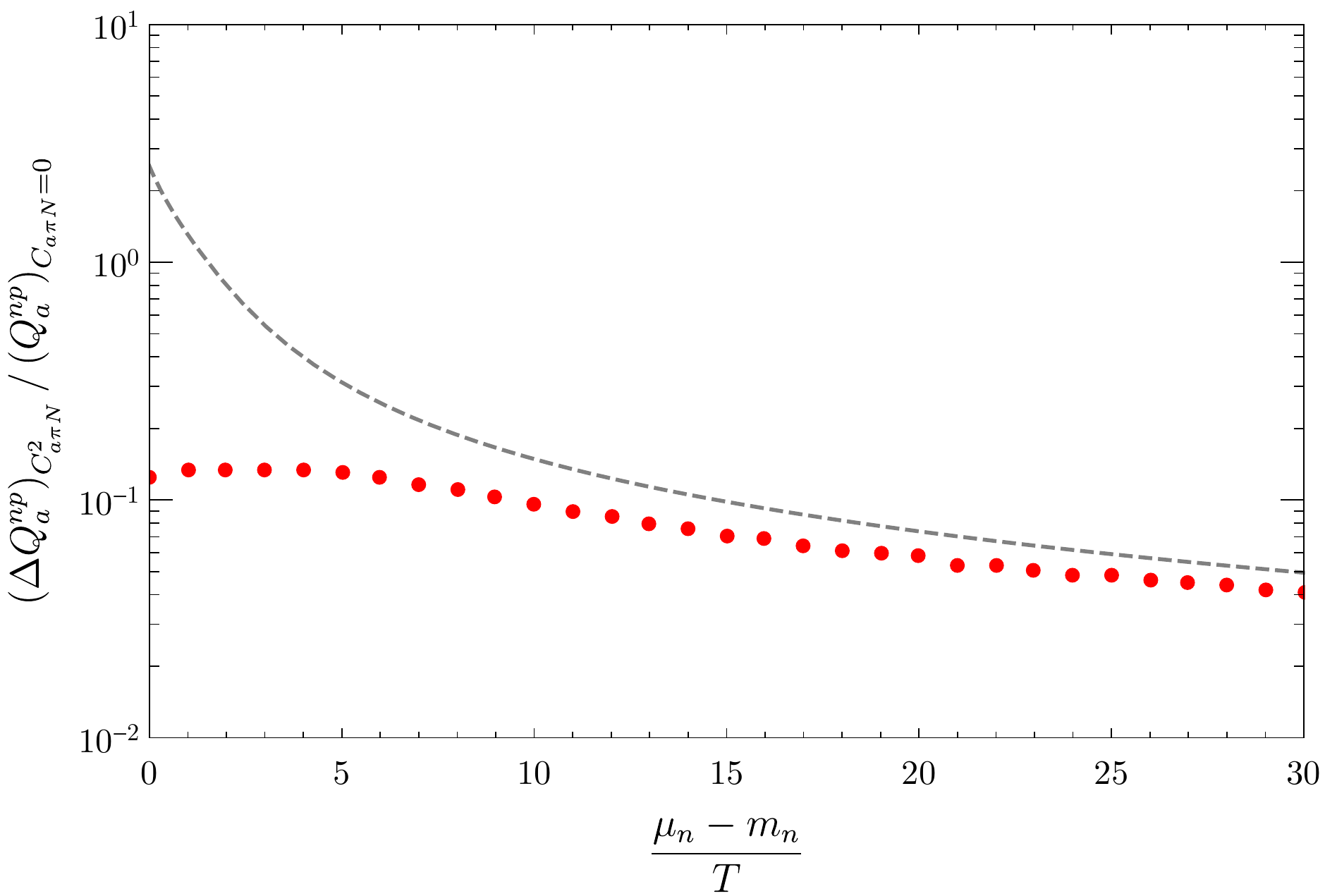} 
  \end{center}
  \caption{The ratio of the axion emissivity contributed from the contact interaction, i.e.~$C_{a\pi N}$ coupling, to that from the other couplings, $C_{ap}$ and $C_{an}$. We compare the numerical result (red dots) with an analytic estimation (dashed). For benchmark parameters, we assume $\mu_p-m_p\simeq 0.25(\mu_n-m_n)$ and $C_{ap}=-C_{an}=1/2$.}
\label{diagram:npCapiNratio}
\end{figure}

In the degenerate regime, the axion emissivity for the nucleon-nucleon bremsstrahlung can be derived in an analytical way because the phase space is highly constrained \cite{Iwamoto:1992jp}. 
We can apply the method to include the contribution from the contact term, which is a higher order in the expansion with respect to $|\mathbf{p}_a|/|\mathbf{p}_N|\sim \sqrt{T / m_N}$ (see Eq.~\eqref{np:matrixelem}). Although our system belongs to the non-degenerate regime, the analytical estimation provides some insights for the numerical results. Moreover, the analytical result is applicable for the degenerate system like a neutron star. 
\begin{align} 
\frac{(\Delta Q_a^{np})_{C_{a\pi N}^2}}{(Q_a^{np})_{C_{a\pi N}=0}}
& =
\frac{123\pi^2}{124g_A^2}\frac{T^2}{p_{F, n}^2}\frac{y}{x^2}
C_{a\pi N}^2 \bigg[H\left(\frac{2 x y}{x+y}\right)-H\left(\frac{2 x y}{y-x}\right)\bigg]
\nonumber\\
&\hspace{0.5cm}\times\bigg[
\left(C_-^2+C_+^2\right)F(y)-2\left(C_-^2+C_+^2\right)G(y)
\nonumber\\
&
\hspace{1.2cm}
+
\left(C_-^2+2C_+^2\right)
\left\{
\left(
1-\frac{y}{x} \right)F\left(\frac{2 x y}{y-x}\right)
+\left(1+\frac{y}{x}\right)
F\left(\frac{2 x y}{x+y}\right)
\right\}
\bigg]^{-1}
,
\nonumber
\\
&\simeq
\frac{C_{a\pi N}^2}{C_-^2+3C_+^2}\left(\frac{7.0\,T}{p_{F, n}}\right)^2
,
\nonumber
\\
&=
\frac{\left(C_{ap}-C_{an}\right)^2}{\left(C_{ap}-C_{an}\right)^2+3\left(C_{ap}+C_{an}\right)^2}\left(\frac{7.7\,T}{p_{F, n}}\right)^2
,
\label{eq:emissivitynpratio}
\end{align}
where 
$x
\equiv m_\pi/(2p_{F, n})$,
$
y
\equiv  m_\pi/(2p_{F, p}),
$
and
\begin{align}
F(u) 
&\equiv \frac{1}{2}\left(3-\frac{1}{1+u^2}-3 u \cot^{-1}u\right)
,\nonumber
\\
G(v) 
&
\equiv
1-v \cot^{-1} v
,
\nonumber
\\
H(w)
&
\equiv
-\frac{w}{1+w^2}+\cot^{-1} w
.\nonumber
\end{align}
For the second equality of Eq.~(\ref{eq:emissivitynpratio}), we keep the leading order with respect to $x$ and $y$.
Given the parameter values in the main text, e.g.~$T\simeq 40\,{\rm MeV}$ and $p_{F,n}\simeq 200-300\,{\rm MeV}$, Eq.~\eqref{eq:emissivitynpratio} leads to the wrong interpretation that the new contribution from the contact interaction is comparable to the other contributions from the $C_{ap}$ and $C_{an}$ couplings.
The discrepancy between the analytic estimation in Eq.~\eqref{eq:emissivitynpratio} and the numerical results in Eq.~\eqref{fit:np} is caused by applying the formula derived in the degenerate limit to the non-degenerate system \cite{Carenza:2020cis}.
Fig.~\ref{diagram:npCapiNratio} shows the emissivity ratio of the $C_{a\pi N}^2$ contribution with respect to the other contributions from the axion-nucleon couplings for $\mu_n>m_n$. 
Actually, the chemical potential for the proton in the supernova environments \cite{Fore:2019wib} corresponds to the case of $\mu_p< m_p$. 
In order to highlight the difference between the emissivity calculated in degenerated limit and that for non-degenerate case, 
the proton chemical potential is taken to have a simple relation as $\mu_p-m_p\simeq 0.25(\mu_n-m_n)$ and we choose $C_{ap}=-C_{an}=1/2$.
The analytic estimation for the emissivity ratio is shown in Fig.~\ref{diagram:npCapiNratio} as a dashed line.
 Fig.~\ref{diagram:npCapiNratio} gives information on the validity of the degenerate limit.

\bibliographystyle{utphys}
\bibliography{ref}

\end{document}